\documentclass[reprint,twocolumn,superscriptaddress,showpacs,showkeys,preprintnum bers,amsmath,amssymb,floatfix,aps,pra,longbibliography]{revtex4-1}

\usepackage{graphicx}
\usepackage{bm}

\usepackage{placeins}
\usepackage{graphics}
\usepackage{color}
\usepackage{hyperref}
\usepackage{multirow}
\usepackage{blindtext}

\begin{document}


\title{Proximity spin-orbit and exchange coupling in ABA and ABC \\trilayer graphene van der Waals heterostructures
}

\author{Klaus Zollner}
	\email{klaus.zollner@physik.uni-regensburg.de}
	\affiliation{Institute for Theoretical Physics, University of Regensburg, 93053 Regensburg, Germany}
\author{Martin Gmitra}
	\affiliation{Institute of Physics, Pavol Jozef \v{S}af\'{a}rik University in Ko\v{s}ice, 04001 Ko\v{s}ice, Slovakia}
\author{Jaroslav Fabian}
	\affiliation{Institute for Theoretical Physics, University of Regensburg, 93053 Regensburg, Germany}
\date{\today}

\begin{abstract}
We investigate the proximity spin-orbit and exchange couplings in ABA and ABC trilayer graphene
encapsulated within monolayers of semiconducting transition-metal dichalcogenides and 
the ferromagnetic semiconductor Cr$_2$Ge$_2$Te$_6$.
Employing first-principles calculations we obtain the electronic structures of the multilayer stacks and extract 
the relevant proximity-induced orbital and spin interaction parameters by fitting the low-energy bands to model Hamiltonians. We also demonstrate the tunability of the proximity effects by a transverse  electric field. Using the model Hamiltonians we also study mixed spin-orbit/exchange coupling encapsulation, which allows to tailor the spin interactions very efficiently by the applied field. We also summarize the spin-orbit physics of bare ABA, ABC, and ABB trilayers, and provide, along with the first-principles results of the electronic band structures, density of states, spin splittings, and electric-field tunabilities of the bands, qualitative understanding of the observed behavior and realistic model parameters as a resource for model simulations of transport and correlation physics in trilayer graphene. 
\end{abstract}

\maketitle

\section{Introduction}

Two-dimensional (2D) van der Waals (vdW)
materials are vital building blocks in the design of ultracompact electronic and spintronic devices \cite{Han2014:NN,Fabian2007:APS,Gong2019:SC,Li2019:AM,Cortie2019:AFM}.
In this context, proximity-induced phenomena \cite{Zutic2019:MT,Sierra2021:NN}
were found to be of great importance, since 2D materials  influence each others electronic and spin properties in vdW heterostructures. 
Recently, it has been demonstrated that superconductivity \cite{Li2020:PRB,Moriya2020:PRB,Trainer2020:ACS}, magnetism \cite{Zollner2016:PRB,Zollner2018:NJP,Zollner2019a:PRB,Zollner2020:PRB,Zhang2015:PRB,Zhang2018:PRB,Vila2021:arxiv,Wang2015:PRL,Xu2018:PRB,Yang2013:PRL,Zhong2020:NN,Averyanov2018:ACS,Ciorciaro2020:arxiv}, and spin-orbit coupling (SOC) \cite{Gmitra2015:PRB,Gmitra2016:PRB,Zollner2019b:PRB,Zollner2021:PSSB,Song2018:NL,Zhang2014:PRL,Avsar2017:ACS,Avsar2014:NC,Alsharari2018:PRB,Frank2016:PRB}, can be induced on demand in such weakly glued vdW multilayers. In addition, gating, twisting, stacking, and straining are efficient tunability knobs to tailor these spin interactions \cite{Zollner2019a:PRB,Song2018:NL,David2019:arxiv,Avsar2017:ACS,Ghiasi2019:NL,Benitez2020:NM,Luo2017:NL,Safeer2019:NL,Herlin2020:APL,Zollner2021:arXiv2,Naimer2021:arxiv}, while the individual materials also preserve a great degree of autonomy. 

In twisted vdW heterostructures  \cite{Carr2017:PRB,Hennighausen2021:ES,Ribeiro2018:SC,Carr2020:NRM} the interlayer interaction can be controlled by the twist angle. In this context, bilayer graphene (BLG) was the first model playground for gate- and twist-tunable correlated physics \cite{Cao2018:Nat,Cao2018a:Nat,Arora2020:arxiv,Stepanov2020:Nat,Lu2019:Nat,Sharpe2019:SC,Ribeiro2018:SC}, as well as for layer-dependent proximity-induced spin interactions \cite{Gmitra2017:PRL, Amann2021:arxiv, Zollner2020:PRL,Lin2021:arxiv,Wang2019:NL,Zollner2018:NJP, Island2019:Nat, Cardoso2018:PRL,Tiwari2021:PRL,Zollner2021:arXiv}. 
At very small twist angles ($\approx 1^{\circ}$), a sharp peak arises in the density of states (DOS) of twisted BLG, which is associated with flat bands \cite{Bistritzer2011:PNAS} in the dispersion. By doping the twisted BLG, it can become insulating, ferromagnetic \cite{Sharpe2019:SC}, or superconducting \cite{Cao2018:Nat,Cao2018a:Nat,Arora2020:arxiv,Stepanov2020:Nat,Lu2019:Nat,Nimbalkar2020:NML,Saito2020:NP}.
As a logical next step, also twisted BLG-BLG structures were considered \cite{Shen2020:NP,Liu2020:Nat,Cao2020:Nat,Burg2019:PRL}, where the flat bands are additionally tunable by a gate field. 

In addition, recent experiments \cite{Lin2021:arxiv} could demonstrate that the proximity-induced Rashba and valley Zeeman SOC in twisted-BLG/WSe$_2$ heterostructures, induces orbital magnetism without the need for a rotational alignment to a hexagonal boron-nitride substrate~\cite{Sharpe2019:SC,Serlin2020:S}. Moreover, the transition-metal dichalcogenide (TMDC) WSe$_2$ can help to stabilize superconductivity in twisted BLG \cite{Arora2020:arxiv}, emphasizing the role of the dielectric environment. For the interpretation of such experimental results, it is also important to have qualitative and quantitative knowledge about the proximity effects in TMDC/BLG heterostructures \cite{Zollner2021:arXiv}.

Unlike mono- and bilayer graphene, trilayer graphene (TLG) has not yet been systematically investigated for spin proximity effects. The spin-orbit physics of pristine TLG in ABA and ABC stackings was investigated by ab-initio calculations  \cite{Konschuh2011:Diss} and tight-binding modeling \cite{Kormanyos2013:PRB}, revealing many subtleties of the low-energy bands stemming from the presence of $d$-orbitals which give rise to spin-orbit splittings on the order of 10~$\mu$eV. 
Depending on the stacking order of the three layers, either ABA or ABC, very different electronic structures can be realized with distinct features in electronic and spin transport \cite{Jhang2011:PRB,Ghosh2012:JAP} and gate tunable SOC \cite{Chen2012:NL}.
TLG is also important from the topological perspective, since the ABC structure potentially hosts quantum spin Hall and quantum valley Hall states \cite{Rehman2017:CPB}, while there is evidence for a giant topological magnetic moment in ABA TLG \cite{Ge2021:arxiv}. Also interesting is the energetics of the different stackings, see, for example, Ref. \cite{Guerrero2021:arxiv}

More recently, TLG has emerged as a novel platform for correlated electrons \cite{Zhu2020:PRL,Park2021:Nat,Chen2019:NP,Chen2020:Nat,Chen2019:Nat,Zhou2021:arxiv,Polshyn2020:Nat}, as there is one more independent layer, leading to a wider range of magic angles. 
As a consequence, the tunability of the electronic and superconducting properties are superior to the ones in BLG, as recently demonstrated 
\cite{Park2021:Nat,Phong2021:arxiv,Zhou2021:arxiv2,Qin2021:arxiv,Cao2021:Nat} and theoretically explained \cite{Chou2021:arxiv}.
Remarkably, a zero-field superconducting diode, signalling an interplay of spin-orbit physics and time-reversal symmetry broken phase, has been demonstrated in twisted-TLG/WSe$_2$ heterostructures \cite{Scheurer2021:arxiv1,Scheurer2021:arxiv2,Scheurer2021:arxiv3}.
For the interpretation and quantitative theoretical understanding of such experiments it is important to have a deeper microscopic knowledge of the spin proximity effects in TLG. Providing such a resource is the goal of our manuscript. 

In particular, we investigate, by performing first-principles calculations,
the electronic structures of ABA and ABC TLG encapsulated within strong SOC semiconductors MoSe$_2$ and WSe$_2$ and 
ferromagnetic monolayers Cr$_2$Ge$_2$Te$_6$ (CGT). We provide the essential band structure information, low-energy band dispersions, spin splittings of the low-energy bands, spin textures, and the behavior of the Dirac bands in the presence of a transverse electric field. To provide reference for the encapsulated systems, we summarize the essential low energy physics of bare ABA and ABC TLG, also including ABB TLG for completeness. While pristine graphene exhibits intrinsic SOC of 10--20~$\mu$eV, encapsulated TLG displays
spin splittings of some meV, strongly tunable by the electric field.

In terms of modeling, we focus on the proximity-induced spin-orbit and exchange couplings, and demonstrate that in order to reproduce the density functional theory (DFT) data, it is sufficient to modify only the spin interactions of the outer layers of encapsulated TLG. The effective low-energy Hamiltonians, which are fitted to the TLG dispersions to obtain reasonable parameter sets, nicely reproduce the first-principles data. Moreover, we show that the model reproduces the data also in the presence of a transverse electric field, whereby modifying only a few effective couplings is sufficient for a quantitative comparison.

Employing the model, with fitted proximity spin-orbit and exchange couplings separately, we then provide predictions for ex-so-tic heterostructures \cite{Zollner2020:PRL,exsotic} which comprise
both a strong SOC monolayer and a ferromagnetic monolayer, encapsulating ABA and ABC TLG. The interplay between spin-orbit and exchange coupling is imprinted onto the corresponding low-energy band structures and most markedly
pronounced by the different band dispersions at $K$ and $K^{\prime}$. We also show that the band dispersions respond sensitively to the applied electric field, which can serve as a knob to change the spin polarization of the low-energy states in ABC TLG.  

In all the studied cases we provide both qualitative understanding, based on the atomic arrangement and hybridization across the layers, and quantitative modeling
with realistic parameters fitted to the DFT results, which should be useful for model simulations of TLG.

The paper is organized as follows: In section~\ref{baretlg}, we first discuss the electronic properties of bare TLG with ABA, ABC, and ABB stackings, under the influence of SOC and an external electric field, to have reference results for the encapsulated structures. In section~\ref{encaptlg}, we present the encapsulated TLG geometries, address the structural setup, and summarize the calculation details  for obtaining the electronic structures. In section~\ref{model}, we introduce the model Hamiltonians that capture the low energy physics of encapsulated TLG (including orbital, SOC, and exchange terms), which are used to fit the DFT-calculated dispersions. In section~\ref{bandfits}, we then show and discuss the DFT-calculated electronic structures, along with the model Hamiltonian fits, of TMDC and CGT encapsulated TLG. 
In section~\ref{efield}, we turn to the electric field tunability of the relevant low energy bands of TMDC encapsulated TLG.
In section~\ref{spininter}, we discuss the interplay of spin interactions in CGT/TLG/TMDC heterostructures,
on the model level, based on our results for TMDC and CGT encapsulated TLG.
Finally, in section~\ref{summ} we conclude the manuscript.

\section{Bare trilayer graphene}
\label{baretlg}

Recent investigations on twisted-graphene/BLG heterostructures, that contain all three high-symmetry TLG stacking types (ABA, ABC, and ABB), reveal highly interesting topological properties \cite{Polshyn2020:Nat,Rademaker2020:PRR,Park2020:PRB,Ma2021:SB}. For example, in Ref.~\cite{Polshyn2020:Nat} the electrical control of magnetism, which arises due to strong correlations by twisting graphene on top of BLG, has been demonstrated.

To have reference results for evaluating the proximity effects in 
encapsulated graphene trilayers, we briefly review the essential electronic properties of bare ABA, ABC, and ABB stackings, focusing on the energy dispersions of the lowest energy bands and the spin-orbit splittings.
The electronic band structures of the bare trilayers were obtained by performing first-principles calculations employing a full potential linearized augmented plane wave (FLAPW) code based on density functional theory (DFT), as implemented in WIEN2k \cite{Wien2k}. This code was already used to calculate SOC in monolayer and bilayer graphene, \cite{Gmitra2009:PRB,Konschuh2010:PRB,Konschuh2012:PRB}, yielding 
results consistent with recent experimental findings \cite{Sichau2019:PRL,Banszerus2020:PRL}.

In our implementation, exchange-correlation effects are treated with the generalized-gradient approximation (GGA) \cite{Perdew1996:PRL}, including the DFT-D3 vdW dispersion correction \cite{Grimme2010:JCP}. We used a very dense $k$-point grid of $120\times 120\times 1$ to accurately determine the Fermi level. The Muffin-tin radius of carbon atoms is $R_{\textrm{C}}=1.34$ and the plane wave cutoff parameter $RK_{\textrm{MAX}}=9.5$. For all TLG stackings, we use the lattice parameter of bare graphene $a = 2.46$~\AA, with fixed interlayer distances of $d = 3.3$~\AA. 
In order to avoid interactions between periodic images of our slab geometries, we add a vacuum of about $20$~\AA~in the $z$ direction. An earlier account of WIEN2k TLG calculations can be found in Ref. \cite{Konschuh2011:Diss}, while multi-orbital tight-binding modeling of ABC TLG in the presence of SOC was performed in Ref. \cite{Kormanyos2013:PRB}.

\subsection{ABA}

\begin{figure*}[htb]
	\includegraphics[width=0.99\textwidth]{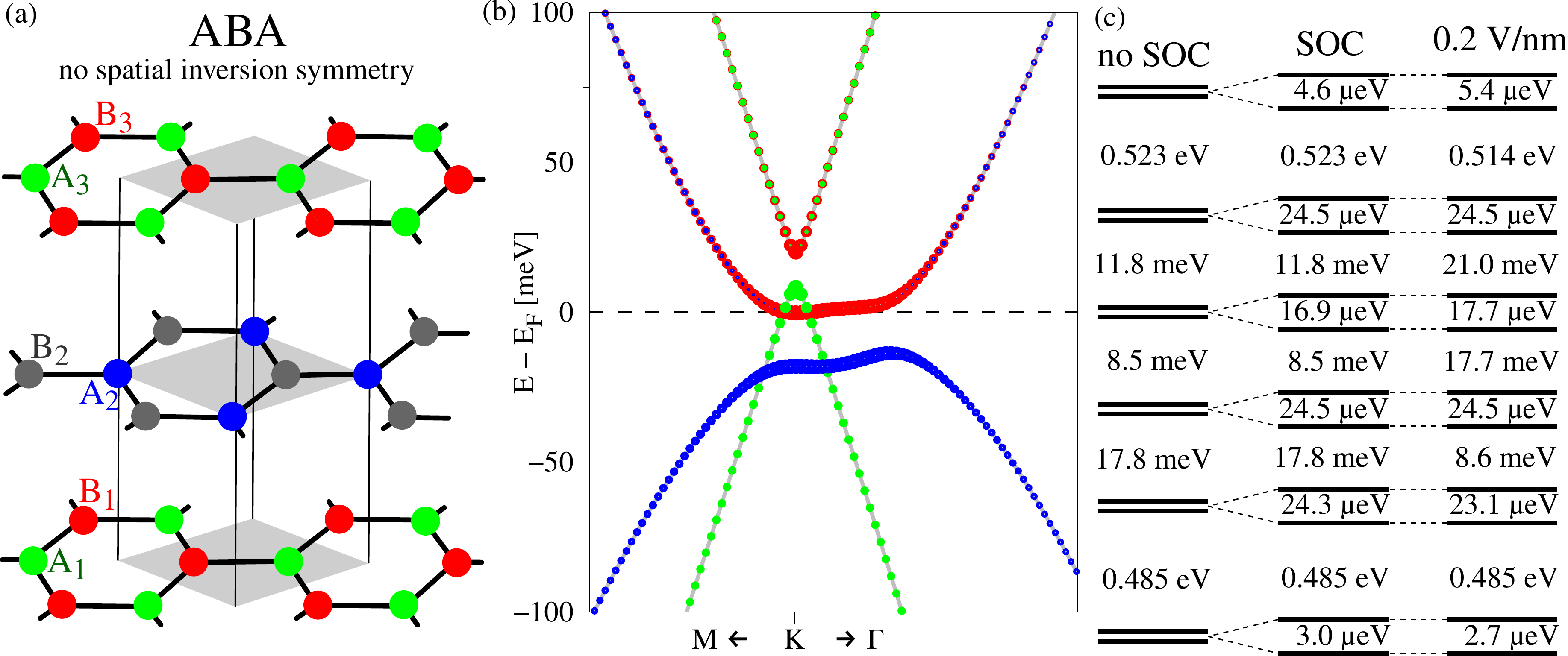}
	\caption{(a) Lattice structure of the ABA TLG. Due to $z$-mirror symmetry, the sublattice atoms with the same color (red, blue, green, and grey) belong together. Grey rhombuses define the unit cells of the layers. 
	(b) The DFT-calculated band structure in the vicinity of the $K$ point. The bands are color-coded by their atomic projections as defined in (a). 
	(c) The energy level diagram of states at the $K$ point without SOC, with SOC, and with SOC in the presence of a perpendicular electric field of 0.2~V/nm. 
 \label{Fig:bare_ABA}}
\end{figure*}

In Fig.~\ref{Fig:bare_ABA} we show the lattice structure, the calculated energy dispersion, and a schematic energy level diagram for ABA TLG.
Due to $z$-mirror symmetry, four nonequivalent atoms are present in the geometry, see Fig.~\ref{Fig:bare_ABA}(a).
The low-energy band structure, see Fig.~\ref{Fig:bare_ABA}(b), shows graphene-like Dirac states, as well as BLG-like parabolic bands, consistent with literature \cite{Partoens2006:PRB,Latil2006:PRL,Koshino2009:PRB2,Aoki2007:SSC}.
The Dirac bands are exclusively formed by orbitals of the atoms in the outermost graphene layers. In contrast, the parabolic bands are formed by atoms B$_1$, A$_2$, and B$_3$, which form non-dimer interlayer pairs.

Without SOC, the bands remain spin-degenerate, see the scheme in Fig.~\ref{Fig:bare_ABA}(c). Including SOC, the degeneracy is lifted due to the absence of space-inversion symmetry and the low-energy bands are split by about 20~$\mu$eV (corresponding to the intrinsic SOC of pristine graphene monolayers \cite{Gmitra2009:PRB}).
The high-energy states --- not shown in Fig.~\ref{Fig:bare_ABA}(b) but included in the energy level diagram in Fig.~\ref{Fig:bare_ABA}(c) --- which are formed by atoms A$_1$, B$_2$, and A$_3$, are about 500~meV away from the Dirac point at the Fermi level. The reason is that these atoms are coupled by direct interlayer hopping, pushing the corresponding bands away from the Fermi level, similar to bare BLG \cite{Konschuh2012:PRB}. 
The spin-orbit splittings of the high-energy bands are just a few $\mu$eV. The diagram in Fig.~\ref{Fig:bare_ABA}(c) also indicates the effect of an applied transverse electric field which introduces a potential difference between the outermost layers. The
field essentially reshuffles the energy levels, but mainly the band offsets of the Dirac and parabolic low-energy bands change, while the spin-orbit splittings remain nearly unchanged for the electric field as large as 0.2~V/nm. 
In appendix \ref{AppA}, we show and discuss the evolution of the ABA TLG dispersion and DOS for electric fields up to 1~V/nm. We also provide model Hamiltonian fit results there, that nicely reproduce the dispersion.

\subsection{ABC}

\begin{figure*}[htb]
	\includegraphics[width=0.99\textwidth]{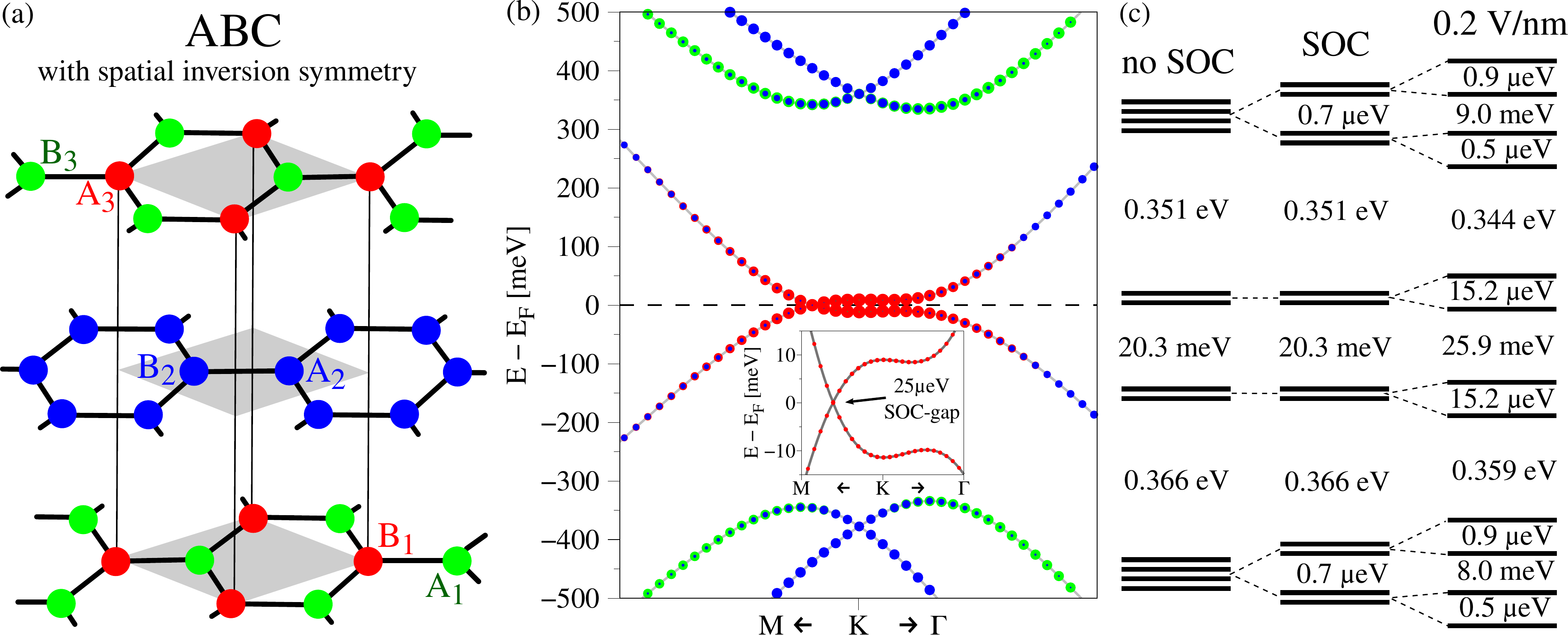}
	\caption{(a) Lattice structure of the ABC TLG. Due to space inversion symmetry, the sublattice atoms of the same color (red, blue, and green) contribute equally to the energy bands. Grey rhombuses define the unit cells of the layers. 
	(b) The DFT-calculated band structure in the vicinity of the $K$ point. The bands are color-coded by their atomic projections as defined in (a). 
	The inset in (b) shows a zoom to the low-energy bands near the Fermi level, indicating the cone-like touching point, where a gap of about 25~$\mu$eV arises due to SOC. 
	(c) The energy level diagram of the low and high-energy states at the $K$ point without SOC, with SOC, and with SOC in the presence of a perpendicular electric field of 0.2~V/nm. 
 \label{Fig:bare_ABC}}
\end{figure*}

In Fig.~\ref{Fig:bare_ABC} we show the lattice structure, the calculated energy dispersion, and an energy-level diagram for ABC TLG.
This structure has spatial inversion symmetry and only 
three nonequivalent atoms are present in the geometry, see Fig.~\ref{Fig:bare_ABC}(a).
From the band structure at the $K$ point, see Fig.~\ref{Fig:bare_ABC}(b), we find two flat low-energy bands which are formed by the outer-layer atoms B$_1$ and A$_3$, although A$_2$ and B$_2$ become prominent at larger momenta away from $K$. The four high-energy bands, which are split off from the Fermi level by more than 0.35~eV, are built by sublattices A$_1$, A$_2$, B$_2$, and B$_3$. 
The presented dispersion is consistent with earlier reports \cite{Latil2006:PRL,Koshino2009:PRB,Aoki2007:SSC,Zhang2010:PRB,Lu2006:APL}.

At the $K$ point, the low-energy bands are twofold spin degenerate, while the high-energy bands are even fourfold degenerate, in the absence of SOC, see Fig.~\ref{Fig:bare_ABC}(c). 
Due to spatial inversion symmetry, all bands remain spin degenerate, even when SOC is turned on. However, a tiny (about 0.7 $\mu$eV) spin-orbit gap opens at the high-energy levels and removes the fourfold degeneracy.
Zooming into the Fermi level, we find a 
cone-like touching feature of the low-energy bands along the $K \rightarrow M$ high-symmetry line, see Fig.~\ref{Fig:bare_ABC}(b). When SOC is included, the Dirac cones are split by 25~$\mu$eV, which corresponds to the spin-orbit gap, caused almost solely by $d$ orbitals of a pristine graphene monolayer \cite{Gmitra2009:PRB}.

When a perpendicular electric field is applied across the ABC TLG, the inversion symmetry, along with the spin degeneracy of the bands, gets lifted, see Fig.~\ref{Fig:bare_ABC}(c). The field introduces a potential difference between the outermost layers, leading to further gap openings in the spectrum.
By increasing the field amplitude, the low-energy spin splittings at the $K$ point first increase, then saturate at about $25~\mu$eV, and finally decrease again \cite{Konschuh2011:Diss}. 
For a field of 0.2~V/nm, the splittings of the 
low-energy bands at the $K$ point are about $15~\mu$eV, see Fig.~\ref{Fig:bare_ABC}(c). Near the Dirac cone vertex, the band splittings are $25~\mu$eV, again corresponding to the intrinsic SOC of monolayer graphene~\cite{Gmitra2009:PRB}.
Such intrinsic splittings, that get exposed by an applied electric field and that do not depend on the field above some crossover value, are also present in BLG \cite{Konschuh2012:PRB} and lead to a marked spin relaxation anisotropy as already experimentally detected \cite{Leutenantsmeyer2018:PRL}. 

\subsection{ABB}

\begin{figure*}[htb]
	\includegraphics[width=0.99\textwidth]{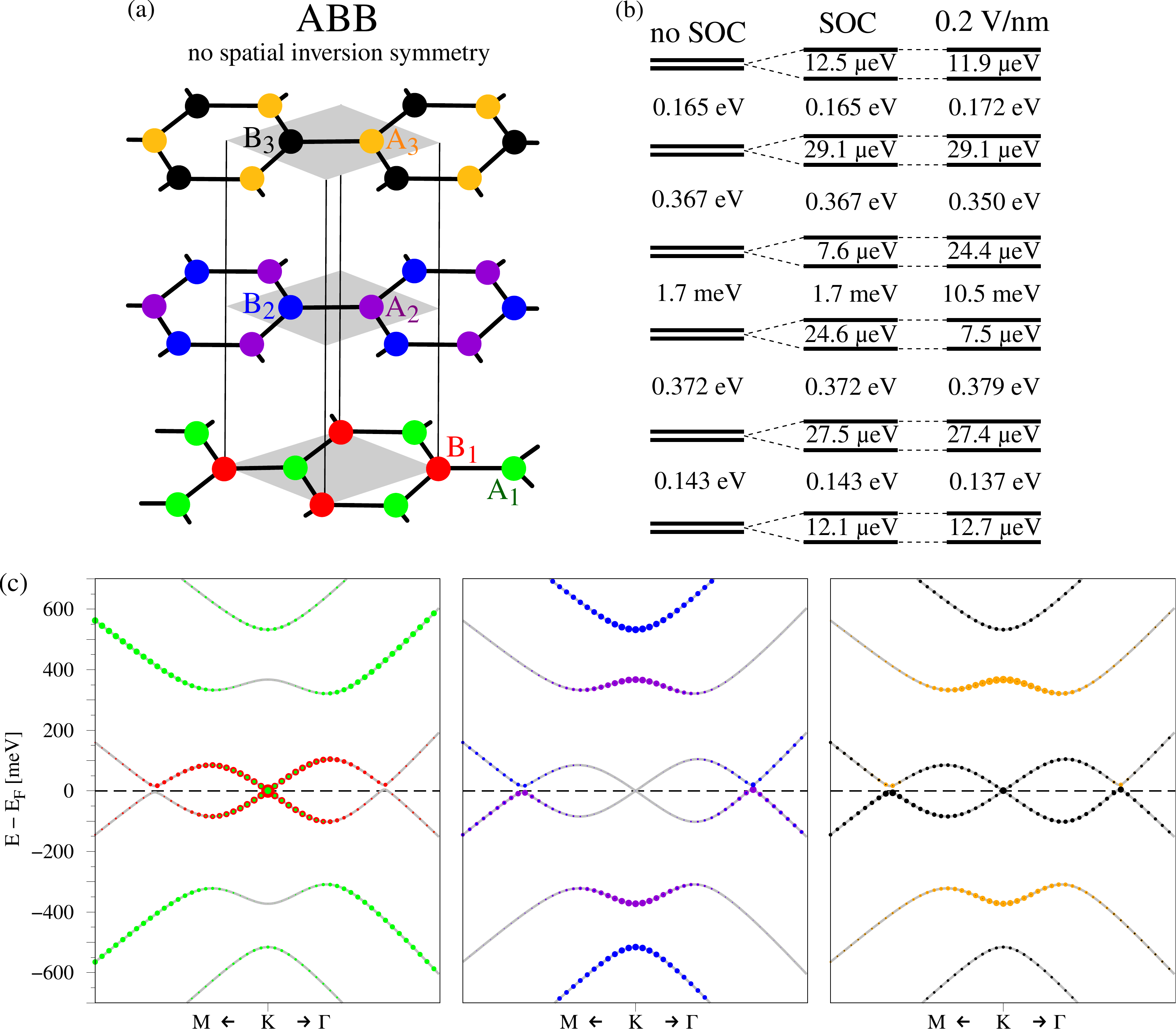}
	\caption{(a) Lattice structure of the ABB TLG. Due to the absence of spatial inversion and $z$-mirror symmetry, all sublattice atoms are different (red, green, blue, purple, black, orange). Grey rhombuses define the unit cells of the layers. 
	(b) The energy level diagram of states at the $K$ point without SOC, with SOC, and with SOC in the presence of a transverse electric field of 0.2~V/nm. 
	(c) The DFT-calculated band structure in the vicinity of the $K$ point without SOC. From left to right, we project onto the different basis atoms from the three graphene layers, with the color code as defined in (a).
 \label{Fig:bare_ABB}}
\end{figure*}

In Fig.~\ref{Fig:bare_ABB} we show the lattice structure, the calculated dispersion, and an energy level diagram for ABB TLG.
Due to the absence of spatial inversion and $z$-mirror symmetry, all sublattice atoms are different, see Fig.~\ref{Fig:bare_ABB}(a).
The band structure, see Fig.~\ref{Fig:bare_ABB}(c), features parabolic high-energy bands at around 600~meV at the $K$ point, predominantly formed by B$_2$ atoms. The reason is that B$_2$ is interlayer-coupled to both surrounding layers, splitting these states off to high energies. 
The low-energy bands near the $K$ point are almost exclusively formed by B$_1$, with a small contribution from A$_1$ and B$_3$. The reason is that B$_1$ does not couple directly to the other layers, so the corresponding band remains close to the Dirac point. 
The intermediate bands are formed by A$_2$ and A$_3$ near the $K$ point, because they are only connected by one interlayer coupling. 
The energy level diagram, see Fig.~\ref{Fig:bare_ABB}(b), shows that the bands remain spin-degenerate without SOC. Including SOC lifts the degeneracy, because spatial inversion symmetry is absent. 
The low-energy valence (conduction) band is split by about 25 $\mu$eV (8 $\mu$eV).
A transverse electric field further opens the gap at the $K$ point, while swapping the spin-orbit splittings---and the orbital composition of the bands themselves---of the low-energy valence and conduction bands (see also Fig.~\ref{Fig:Efield_ABB} in appendix \ref{AppB}). The high-energy bands remain nearly the same, except that band offsets are tunable by the field. 
In Appendix \ref{AppB} we show the evolution of the ABB TLG dispersion and DOS for electric fields up to 1~V/nm. We also provide model Hamiltonian fit results there, that nicely reproduce the dispersion.

For the presented TLG dispersions, we have fixed all the interlayer distances to $d = 3.3$~\AA.
In Appendix \ref{AppC} we compare the dispersions and DOS for ABA, ABC, and ABB TLG with relaxed interlayer distances. Especially the ABB dispersion and DOS are markedly modified, since the interlayer distances become highly asymmetric.

\subsection{Electric Field Effects}

    \begin{figure*}[htb]
     \includegraphics[width=.99\textwidth]{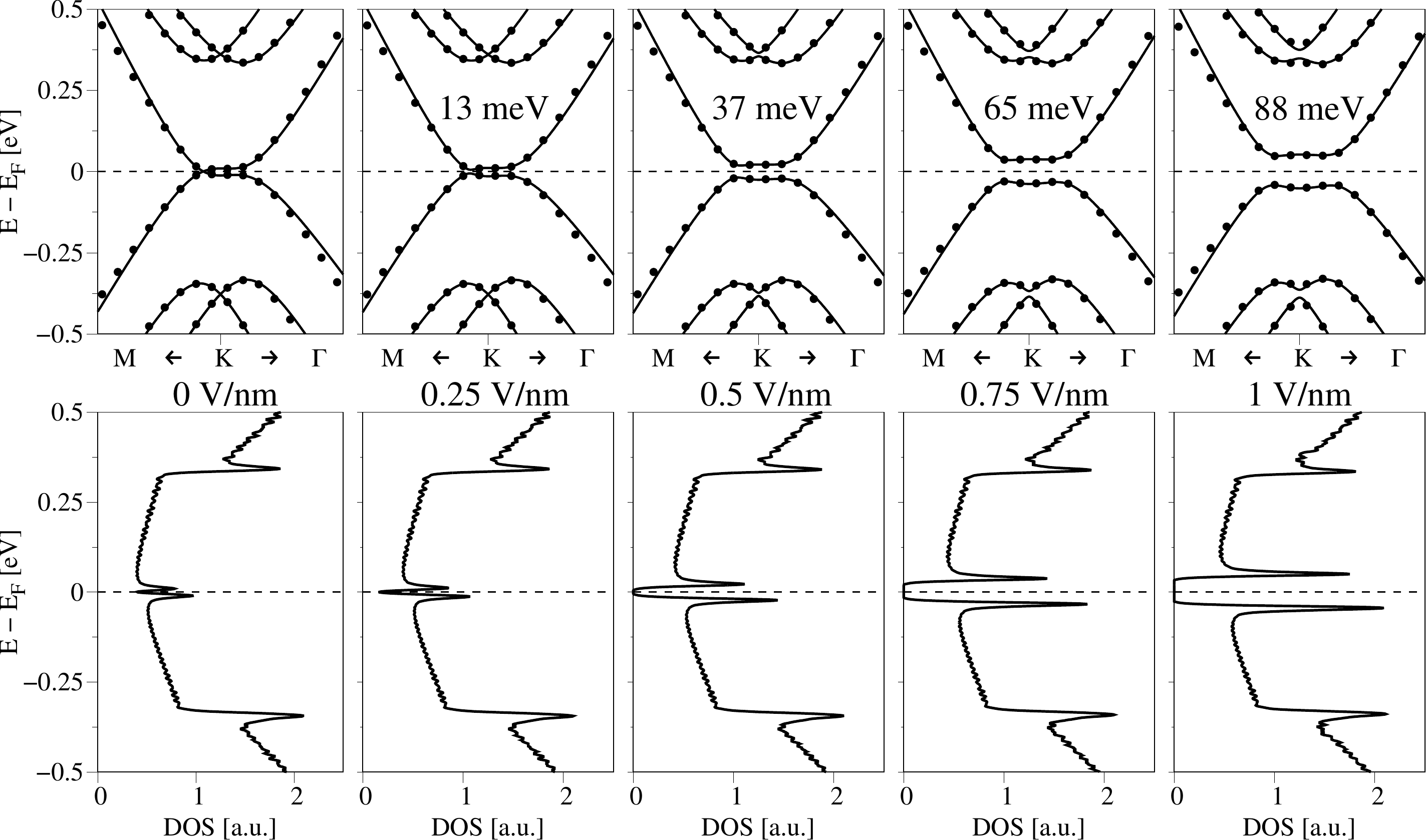}
     \caption{Top: Zooms to the ABC TLG bands in the vicinity of the $K$ point. We compare DFT data (symbols) with model Hamiltonian fits (solid lines) employing parameters from Table~\ref{tab:fit_bare_ABC} in appendix \ref{AppA}. In the dispersion, we specify the band gap at the Fermi level. Bottom: The corresponding calculated density of states (DOS). From left to right, we increase the transverse electric field from 0 to 1~V/nm. 
     }\label{Fig:bands_DOS_bare_ABC}
    \end{figure*}

An important factor for correlation physics in TLG
is the flatness of the dispersion, associated with van Hove singularities (VHS) in the DOS. Recently, superconductivity in ABC TLG was demonstrated \cite{Zhou2021:arxiv2}, which is certainly strongly related with the electric field tunability of these VHS. Below we describe the electric tunability of ABC TLG, while in appendices \ref{AppA} and \ref{AppB} we show and discuss the electric field behavior of ABA and ABB TLG dispersions and their corresponding DOS.

In Fig.~\ref{Fig:bands_DOS_bare_ABC} we show the evolution of the dispersion and DOS for ABC TLG for electric fields up to 1~V/nm.
The field opens a band gap at the Fermi level, since the low-energy bands are formed by atoms B$_1$ and A$_3$ (see Fig.~\ref{Fig:bare_ABC}) of the outermost graphene layers, now located in different potentials. Increasing the field up to 1~V/nm opens a sizable gap of about 90~meV in the spectrum. Also in the high-energy bands, gaps are introduced by the field. However, because they are split-off in energy from the Fermi level, they are not so important for transport experiments. Viewing the corresponding DOS, we find VHS associated with the low-energy bands. As the electric field separates these bands in energy, they further flatten and the VHS become strongly pronounced. 

Given the similarity of the electric field behavior of the low energy bands of BLG \cite{Konschuh2012:PRB} and ABC TLG, one can expect, for example, similar tunability of the valley g-factor as demonstrated for BLG \cite{Moulsdale2020:PRB, Lee2020:PRL}. Moreover, one can also expect a spin-orbit valve operation in ABC TLG, as experimentally and theoretically demonstrated in BLG/TMDC heterostructures \cite{Gmitra2017:PRL,Island2019:Nat}. 

In Fig.~\ref{Fig:bands_DOS_bare_ABC} we also compare DFT data and model Hamiltonian fits. Employing the parameters from Table~\ref{tab:fit_bare_ABC} in appendix \ref{AppA}, we can perfectly fit the dispersions with applied electric field. The model Hamiltonian will be introduced in the following section, when discussing encapsulated TLG heterostructures.

\section{Encapsulated Trilayer Graphene}
\label{encaptlg}
\subsection{Geometry Setup}
\begin{figure*}[!htb]
	\includegraphics[width=0.99\textwidth]{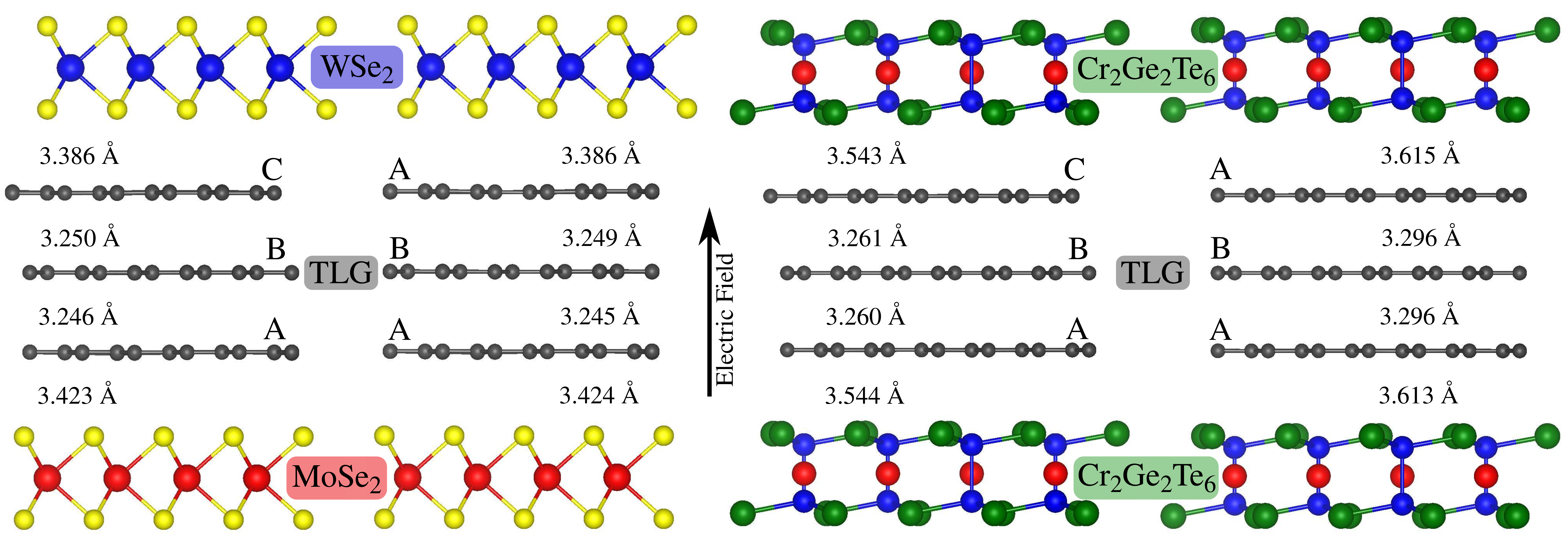}
	\caption{Geometries of the encapsulated TLG heterostructures. Left: TMDC (MoSe$_2$ and WSe$_2$) encapsulated ABA and ABC TLG. Right: CGT encapsulated ABA and ABC TLG. The relaxed interlayer distances are indicated. We also specify the direction of a positive electric field.
 \label{Fig:Structure}}
\end{figure*}

In the following we focus on TLG in ABA and ABC stacking, which are of most current interest. For both stacking types, we consider $4 \times 4$ supercells of TLG to be encapsulated by $3 \times 3$ monolayer supercells of bottom MoSe$_2$ and top WSe$_2$. 
In the case of TMDC encapsulation, we use the lattice constant of pristine TLG ($2.46~\textrm{\AA}$) and
the MoSe$_2$ and WSe$_2$ lattice constants are barely strained (below 0.3\% \cite{Schutte1987:JSSC,James1963:AC}) to $3.28~\textrm{\AA}$. 
Moreover, in the case of TMDC encapsulation, the twist angles between all layers is $0^{\circ}$. Here, we do not investigate different twist angles. 
Note that the electronic structures of TMDCs are very sensitive to strain \cite{Zollner2019:strain}, especially in terms of band gap. In contrast, the Dirac states of graphene are quite robust against biaxial strain~\cite{Si2016:RSC,Choi2010:PRB}.
Moreover, upon twisting, proximity SOC in graphene/TMDC heterostructures can be strongly modified, which has been investigated in Refs.~\cite{David2019:arxiv,Li2019:PRB,Naimer2021:arxiv}.
Especially in Ref.~\cite{Naimer2021:arxiv}, DFT calculations have shown that straining graphene, while leaving the TMDC unstrained, determines the position of the Dirac states within the TMDC band gap. This band offset can be tuned by gating, thereby influencing proximity SOC. We believe that these findings would also apply for our heterostructures of TMDC encapsulated TLG.

In addition, we consider $5 \times 5$ supercells of TLG to be encapsulated within $\sqrt{3} \times \sqrt{3}$ supercells of ferromagnetic CGT. 
In the case of CGT encapsulation, we also keep the lattice constant of pristine TLG but stretch the CGT lattice constant by roughly 4\% from $6.8275~\textrm{\AA}$ \cite{Carteaux1995:JPCM} to $7.1014~\textrm{\AA}$, for better comparability of the encapsulation cases. Note that, similar to the TMDCs, CGT also shows a strain-tunable band gap~\cite{Li2014:JMCC,Chen2015:PRB,Liu2021:N}.
Again, the twist angle is crucial for proximity exchange in graphene/CGT heterostructures \cite{Zollner2021:arXiv2}. 
In our geometries, the relative twist angle between top (bottom) CGT and the top (bottom) graphene layer is $30^{\circ}$.
In Ref.~\cite{Zollner2021:arXiv2}, it has been explicitly shown by DFT calculations, that mainly twisting influences proximity exchange, while straining determines the position of the graphene Dirac states within the CGT band gap. This band offset is again tunable by gating, thereby influencing proximity exchange. Also these findings should be applicable for our TLG structures. 

Initial atomic structures are set up with the atomic simulation environment (ASE) \cite{ASE} and visualized with VESTA software \cite{VESTA}, see Fig.~\ref{Fig:Structure}. 
Our choice of the encapsulating monolayers is based on the fact that TMDCs and CGT are semiconductors, providing strong spin-orbit and exchange couplings to graphene, correspondingly \cite{Gmitra2016:PRB,Gmitra2015:PRB,Zollner2018:NJP,Zollner2020:PRL,Zollner2021:arXiv}. Another important factor for our choice is the lattice matching, such that the different materials can be combined in commensurate supercells for periodic DFT calculations, without straining them beyond reasonable limits.

\subsection{Computational Details}
The electronic structure calculations and structural relaxations of 
the TLG-based vdW heterostructures are performed by DFT \cite{Hohenberg1964:PRB} with \textsc{Quantum ESPRESSO} \cite{Giannozzi2009:JPCM}.
Self-consistent calculations are performed with the $k$-point sampling of 
\mbox{$18\times18\times1$} to get converged results for the proximity-induced exchange and SOC. 
We use an energy cutoff for the charge density of $520$~Ry, and
the kinetic energy cutoff for wavefunctions is $65$~Ry for the scalar relativistic pseudopotentials 
with the projector augmented wave method \cite{Kresse1999:PRB} with the 
Perdew-Burke-Ernzerhof exchange correlation functional \cite{Perdew1996:PRL}.
In the case of CGT encapsulation, 
we perform open shell calculations that provide the 
spin polarized ground state and proximity exchange coupling. In addition,   
a Hubbard parameter of $U = 1$~eV is used for Cr $d$-orbitals, 
similar to recent calculations \cite{Gong2017:Nat, Zollner2018:NJP,Zollner2021:arXiv}.
In the case of TMDC encapsulation, we use the relativistic versions of the pseudopotentials, to capture (proximity) SOC effects. 

For the relaxation of the heterostructures, we add
DFT-D2 vdW corrections  \cite{Grimme2006:JCC,Barone2009:JCC} and use 
quasi-newton algorithm based on trust radius procedure. 
Dipole corrections \cite{Bengtsson1999:PRB} are also included to get 
correct band offsets and internal electric fields.
In order to simulate quasi-2D systems, we add a vacuum of $20$~\AA, 
to avoid interactions between periodic images in our slab geometries.
To determine the interlayer distances, the atoms of TLG and the TMDCs 
are allowed to relax only in their $z$ positions 
(vertical to the layers), and the atoms of CGT are allowed to move in all directions,
until all components of all forces are reduced below $10^{-3}$~[Ry/$a_0$], where $a_0$ is the Bohr radius.
The obtained interlayer distances are summarized in Fig.~\ref{Fig:Structure} and are similar to previous reports \cite{Gmitra2017:PRL, Zollner2018:NJP,Zollner2020:PRL}.
Since we have assumed perfectly aligned individual layers, the full heterostructures still have $C_3$ symmetry after relaxation.

\section{Model Hamiltonians}
\label{model}

Here we present the Hamiltonians that we employ to model the low-energy bands of the (encapsulated) ABA and ABC TLG. The basis states are
$|\textrm{C}_{\textrm{A1}}, s\rangle$,  
$|\textrm{C}_{\textrm{B1}}, s\rangle$, 
$|\textrm{C}_{\textrm{A2}}, s\rangle$, 
$|\textrm{C}_{\textrm{B2}}, s\rangle$, 
$|\textrm{C}_{\textrm{A3}}, s\rangle$, 
$|\textrm{C}_{\textrm{B3}}, s\rangle$, providing
12 eigenvalues $\varepsilon_{1-12}$, for both spin species $s = \{\uparrow,\downarrow\}$ of each C atom. 
In this basis the Hamiltonian comprises several terms \cite{Konschuh2011:Diss,Zhang2010:PRB,Koshino2009:PRB,Kormanyos2013:PRB,Koshino2010:PRB}
\begin{equation}
\mathcal{H} =  \mathcal{H}_{\textrm{orb}} + \mathcal{H}_{\textrm{soc}}+\mathcal{H}_{\textrm{ex}}+E_D.
\end{equation}
The orbital terms for ABA and ABC TLG consist of $p_z$ intra- and interlayer hoppings 
\begin{widetext}
\begin{flalign}
\mathcal{H}_{\textrm{orb}}^{\textrm{ABA}} = & \begin{pmatrix}
\Delta+V_1 & \gamma_0 f(\bm{k}) & \gamma_4 f^{*}(\bm{k}) & \gamma_1 & \gamma_5 & 0 \\
\gamma_0 f^{*}(\bm{k}) & \eta+V_1 & \gamma_3 f(\bm{k}) & \gamma_4 f^{*}(\bm{k}) & 0 & \gamma_2 \\
 \gamma_4 f(\bm{k}) & \gamma_3 f^{*}(\bm{k}) & V_2 & \gamma_0 f(\bm{k}) & \gamma_4 f(\bm{k}) & \gamma_3 f^{*}(\bm{k}) \\
\gamma_1 & \gamma_4 f(\bm{k}) & \gamma_0 f^{*}(\bm{k}) & \Delta+V_2 & \gamma_1 & \gamma_4 f(\bm{k})\\
\gamma_5 & 0 & \gamma_4 f^{*}(\bm{k}) & \gamma_1 & \Delta-V_1 & \gamma_0 f(\bm{k}) \\
0 & \gamma_2 & \gamma_3 f(\bm{k}) & \gamma_4 f^{*}(\bm{k}) & \gamma_0 f^{*}(\bm{k}) & \eta-V_1
\end{pmatrix} \otimes s_0, \\
\mathcal{H}_{\textrm{orb}}^{\textrm{ABC}} = & \begin{pmatrix}
\Delta+V_1 & \gamma_0 f(\bm{k}) & \gamma_4 f^{*}(\bm{k}) & \gamma_1 & 0 & 0 \\
\gamma_0 f^{*}(\bm{k}) & \eta+2V_1 & \gamma_3 f(\bm{k}) & \gamma_4 f^{*}(\bm{k}) & \gamma_6 & 0 \\
 \gamma_4 f(\bm{k}) & \gamma_3 f^{*}(\bm{k}) & \Delta+V_2 & \gamma_0 f(\bm{k}) & \gamma_4 f^{*}(\bm{k}) & \gamma_1 \\
\gamma_1 & \gamma_4 f(\bm{k}) & \gamma_0 f^{*}(\bm{k}) & \Delta+2V_2 & \gamma_3 f(\bm{k}) & \gamma_4 f^{*}(\bm{k})\\
0 & \gamma_6 & \gamma_4 f(\bm{k}) & \gamma_3 f^{*}(\bm{k}) & \eta-V_1 & \gamma_0 f(\bm{k}) \\
0 & 0 & \gamma_1 & \gamma_4 f(\bm{k}) & \gamma_0 f^{*}(\bm{k}) & \Delta-V_1
\end{pmatrix} \otimes s_0.
\end{flalign}
\end{widetext}

The spin-orbit coupling and exchange terms are the same for both stackings in the above basis:

\begin{widetext}
\begin{flalign}
\mathcal{H}_{\textrm{soc}}+\mathcal{H}_{\textrm{ex}}= &
\begin{pmatrix}
(\tau \lambda_{\textrm{I}}^\textrm{A1}-\lambda_{\textrm{ex}}^\textrm{A1})s_z & 2\textrm{i}\lambda_{\textrm{R1}}s_{-}^{\tau} & 0 & 0 & 0 & 0\\
-2\textrm{i}\lambda_{\textrm{R1}}s_{+}^{\tau} & (-\tau \lambda_{\textrm{I}}^\textrm{B1}-\lambda_{\textrm{ex}}^\textrm{B1})s_z & 0 & 0 & 0 & 0 \\
0 & 0 & 0 & 0 & 0 & 0 \\
0 & 0 & 0 & 0 & 0 & 0 \\
0 & 0 & 0 & 0 & (\tau \lambda_{\textrm{I}}^\textrm{A3}-\lambda_{\textrm{ex}}^\textrm{A3})s_z & 2\textrm{i}\lambda_{\textrm{R3}}s_{-}^{\tau} \\
0 & 0 & 0 & 0 & -2\textrm{i}\lambda_{\textrm{R3}}s_{+}^{\tau} & (-\tau \lambda_{\textrm{I}}^\textrm{B3}-\lambda_{\textrm{ex}}^\textrm{B3})s_z \\
\end{pmatrix}.
\end{flalign}
\end{widetext}

\begin{figure}[htb]
	\includegraphics[width=0.99\columnwidth]{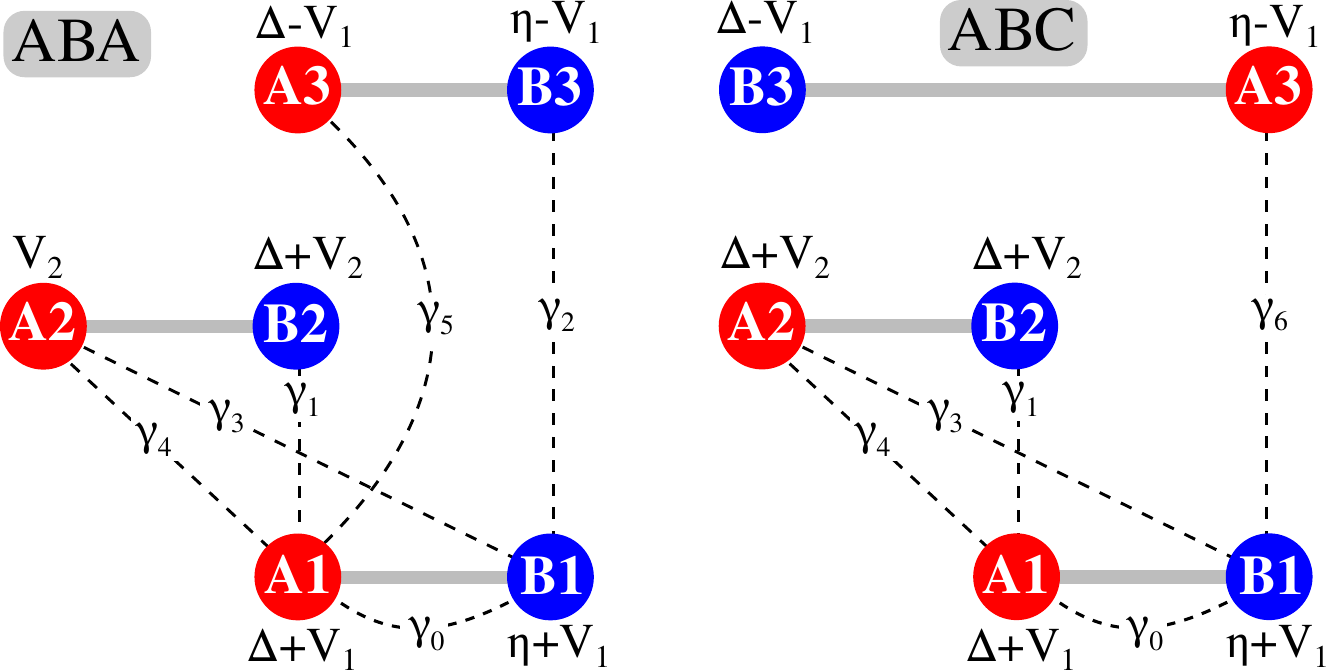}
	\caption{Schematic illustration of the ABA (left) and ABC (right) TLG lattices, showing the relevant intra- and interlayer hoppings $\gamma_j$, $j = \{ 0,1,2,3,4,5,6 \}$ (dashed lines). In addition, the bottom (top) graphene layer is placed in the potential $V_1$ ($-V_1$), while the middle layer is placed in potential $V_2$. The asymmetries $\Delta$ and $\eta$ arise due to vertical hoppings.  
 \label{Fig:TLG_scheme}}
\end{figure}

We use the linearized version of the nearest-neighbor structural function
$f(\bm{k}) = -(\sqrt{3}a/2)(\tau k_x-\textrm{i}k_y)$, with the graphene lattice constant
$a$ and the Cartesian wave vector components $k_x$ and $k_y$ are measured 
from $\pm K$ for the valley indices $\tau = \pm 1$. 
The Pauli spin matrices are $s_i$, with $i = \{ 0,x,y,z \}$, and 
$s_{\pm}^{\tau} = (s_x\pm \textrm{i}\tau s_y)/2$. 
Here, $\gamma_j$, $j = \{ 0,1,2,3,4,5,6 \}$, describe intra- and interlayer hoppings in TLG, see Fig.~\ref{Fig:TLG_scheme} for an overview.  
The coupling $\gamma_0$ is the nearest neighbor intralayer hopping amplitude between A and B sublattice within each graphene layer. 
The parameter $\gamma_1$ describes the direct interlayer hopping between A and B sublattice of adjacent graphene layers.
Similarly, parameters $\gamma_2$, $\gamma_5$, and $\gamma_6$ are also vertical interlayer hoppings, but between the outermost layers in TLG.
In addition, the hoppings $\gamma_3$ and $\gamma_4$ are  interlayer couplings, but not vertical. They connect next nearest neighbor atoms of adjacent graphene layers.
Vertical hoppings in TLG couple only two atoms, hence they appear without structural function in the Hamiltonian, while the other hoppings couple an atom to three corresponding partner atoms, hence they appear linear in momentum. 

\begin{figure*}[htb]
	\includegraphics[width=0.99\textwidth]{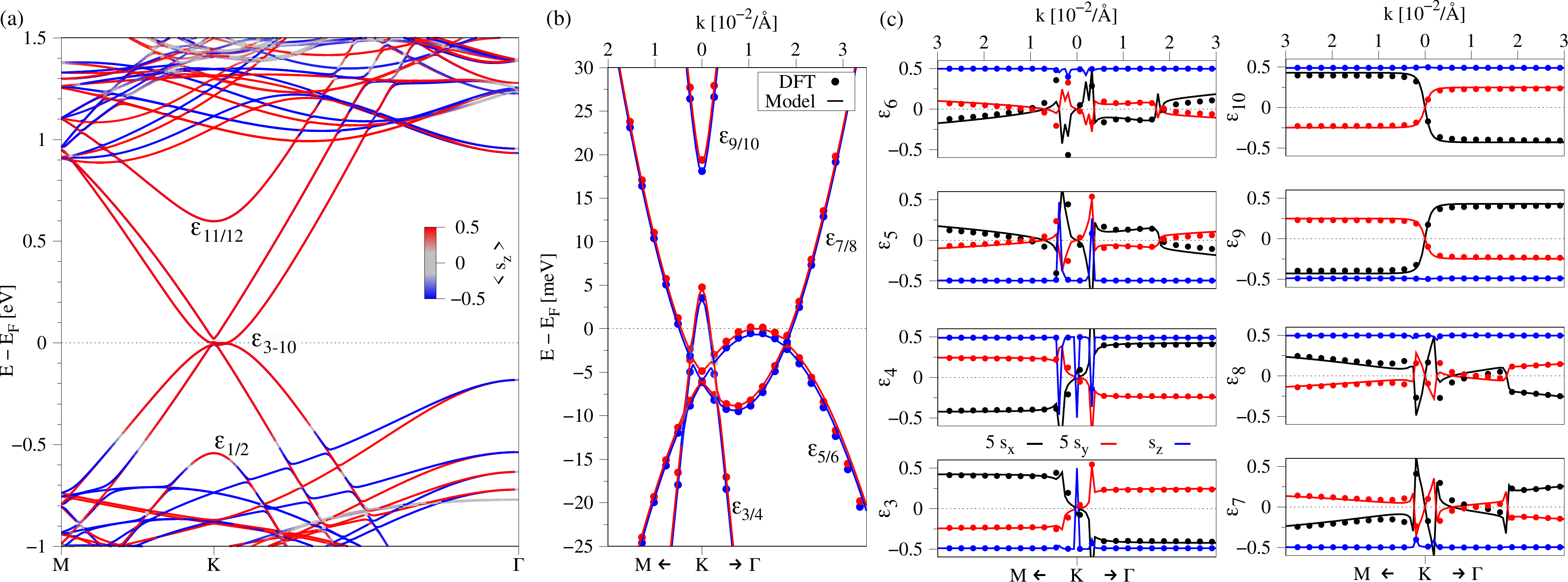}
	\caption{(a) DFT-calculated band structure of the MoSe$_2$/ABA-TLG/WSe$_2$ heterostructure along the $M-K-\Gamma$ path. The color of the bands corresponds to the $s_z$ spin expectation value. We specify the 12 relevant energy bands, $\varepsilon_{1-12}$, corresponding to TLG, which seem to be pairwise spin-degenerate. (b) Zoom to the calculated low-energy bands (symbols) around the $K$ point, corresponding to the band structure in (a), with a fit to the model Hamiltonian (solid lines). The bands are spin-split due to proximity-induced SOC. 
 (c) The spin expectation values of the 8 low-energy bands as labeled in (b).
 The $s_x$ and $s_y$ values are multiplied by a factor of 5 for better visualization. 
 \label{Fig:bands_spins_ABA_TMD}}
\end{figure*}

In general, due to the proximity effect the three layers experience different energy potentials, which we describe
by assigning $V_2$ to the middle layer and $\pm V_1$ to the outer layers. 
In addition, asymmetries $\Delta$ and $\eta$ arise due to the vertical hoppings ($\gamma_1$ and $\gamma_6$ in ABC TLG;  $\gamma_1$, $\gamma_2$, and $\gamma_5$ in ABA TLG).
The combination of parameters $V_1$, $V_2$, 
$\Delta$, and $\eta$ can then describe the on-site energies of the individual sublattices of the different layers within TLG. 
The parameters $\lambda_{\textrm{I}}$ ($\lambda_{\textrm{ex}}$) 
describe the intrinsic or proximity-induced spin-orbit or exchange couplings of the corresponding layer and sublattice atom 
($\textrm{C}_{\textrm{A1}}, \textrm{C}_{\textrm{B1}},  \textrm{C}_{\textrm{A3}}, \textrm{C}_{\textrm{B3}}$).
The parameters $\lambda_{\textrm{R}l}$ represent the Rashba terms of the bottom and top layers $l = \{1,3\}$.
The middle layer, formed by atoms $\textrm{C}_{\textrm{A2}}$ and $\textrm{C}_{\textrm{B2}}$, is far away from the proximitizing TMDC and CGT encapsulation layers. Therefore, we can neglect the corresponding proximity SOC and exchange parameters, as we will see from the band structure fits. 
Actually, in the ABC TLG orbital Hamiltonian, the additional factors of 2 in front of $V_1$ and $V_2$ in the diagonal entries arise due to SOC \cite{Konschuh2011:Diss} and are relevant for the spin-orbit gaps in the high-energy bands, see the energy level diagram in Fig~\ref{Fig:bare_ABC}.
To capture doping effects from the calculations, we introduce another parameter $E_D$, which leads to an energy shift of the model band structure.

Finally, to extract the fit parameters from the DFT, we employ a least-squares routine, taking into account band energies, splittings, and spin expectation values. 
In each case, we first fix the orbital parameters from fitting the dispersion and then fit band splittings and spin expectation values to find the spin-orbit and exchange parameters.

\section{Band Structure and Fit Results}
\label{bandfits}

\subsection{TLG encapsulated within WSe$_2$ and MoSe$_2$}

Monolayer TMDCs are direct band gap semiconductors with strong SOC. Due to proximity of graphene to a TMDC, a significant amount of SOC is introduced in graphene, leading to a splitting of the Dirac states on the order of 1~meV. In the case of TMDC-encapsulated TLG, only the outer graphene layers experience proximity coupling, due to the short rangeness of the proximity effect in vdW heterostructures. This will lead to spin splittings in the TLG band structure, associated with bands originating from the top and bottom layer sublattice atoms. 

\subsubsection{MoSe$_2$/ABA-TLG/WSe$_2$ stacks}

\begin{figure*}[htb]
	\includegraphics[width=0.99\textwidth]{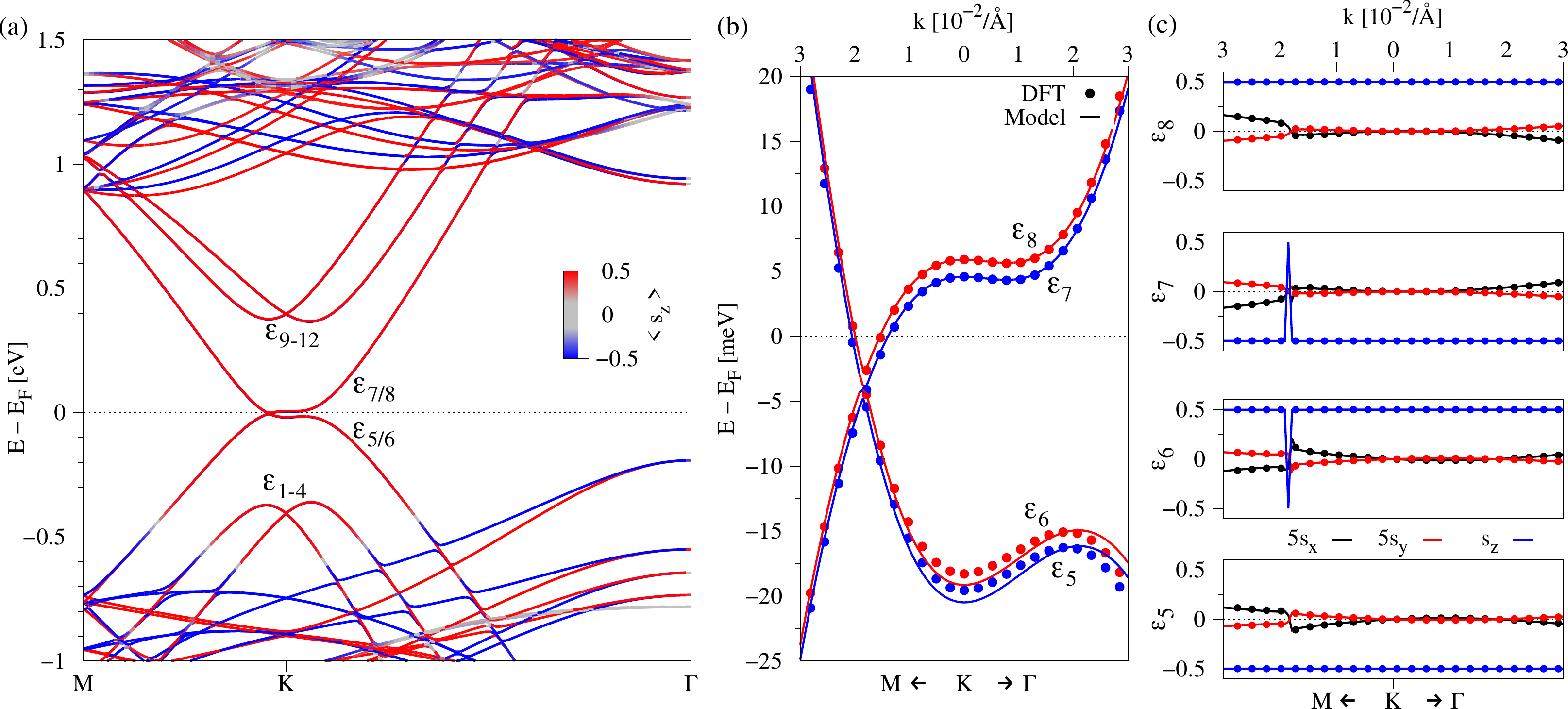}
	\caption{(a) DFT-calculated band structure of the MoSe$_2$/ABC-TLG/WSe$_2$ heterostructure along the $M-K-\Gamma$ path. The color of the bands corresponds to the $s_z$ spin expectation value. We specify the 12 relevant energy bands, $\varepsilon_{1-12}$, corresponding to TLG, which seem to be pairwise spin-degenerate. (b) Zoom to the calculated low-energy bands (symbols) around the $K$ point, corresponding to the band structure in (a), with a fit to the model Hamiltonian (solid lines). The bands are spin-split due to proximity-induced SOC. 
 (c) The spin expectation values of the 4 low-energy bands as labeled in (b). 
  The $s_x$ and $s_y$ values are multiplied by a factor of 5 for better visualization. 
 \label{Fig:bands_spins_ABC_TMD}}
\end{figure*}
\begin{figure}[htb]
	\includegraphics[width=0.99\columnwidth]{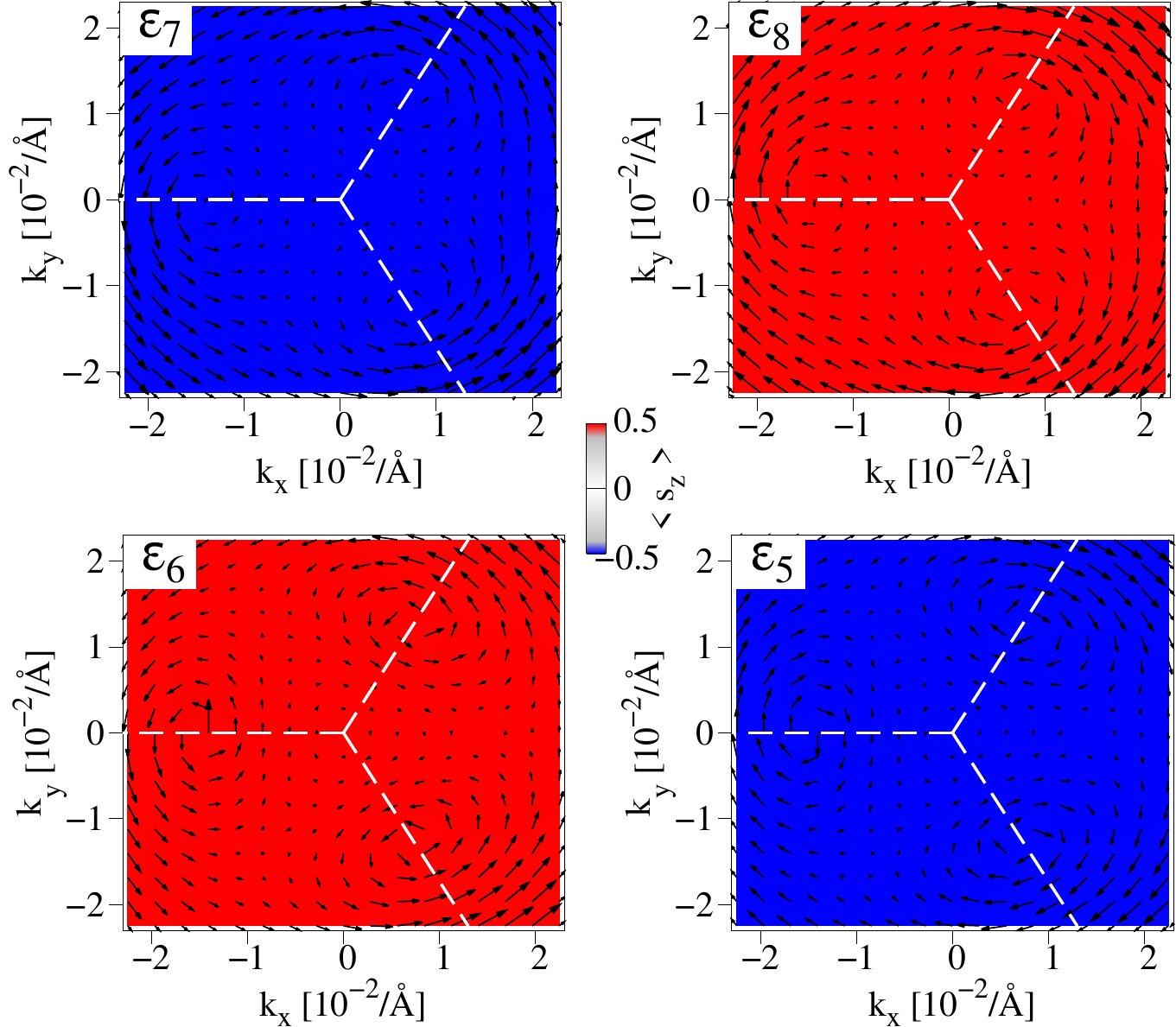}
	\caption{First-principles calculated spin-orbit fields around the $K$ point of the MoSe$_2$/ABC-TLG/WSe$_2$ heterostructure, corresponding to the four low-energy bands $\varepsilon_{5}$ - $\varepsilon_{8}$ in Fig.~\ref{Fig:bands_ABC_Efield} at zero electric field.
 The dashed white lines represent the edges of the hexagonal Brillouin zone with the $K$ point at the center ($k_x = k_y = 0$).
 The $s_z$ spin expectation value determines the color, while in-plane spins are depicted as arrows. Here, we enhanced in-plane spin expectation values by a factor of 10 for better visualization. 
 \label{Fig:spinfield_ABC}}
\end{figure}

We start our discussion by considering the dispersion of the MoSe$_2$/ABA-TLG/WSe$_2$ heterostructure, see Fig.~\ref{Fig:bands_spins_ABA_TMD}(a).
We find that the ABA TLG band structure, featuring the aforementioned 12 energy bands near the $K$ point, is nicely preserved within this heterostructure. Here, we are mainly interested in the 8 low-energy bands, see Fig.~\ref{Fig:bands_spins_ABA_TMD}(b), which resemble a combination of single- and bilayer-graphene low-energy spectra. 
Similar to BLG, the parabolic high-energy bands located at around $\pm 0.5$~eV are mainly formed by states from dimer atoms A$_1$, B$_2$, and A$_3$  (see Fig.~\ref{Fig:pseudospin_ABA} in appendix \ref{AppD}) and are split-off from the Fermi level due to the interlayer hopping $\gamma_1$. The low-energy bands are mainly formed by the non-dimer atoms B$_1$, A$_2$, and B$_3$, but one of them also contains contributions from A$_1$ and A$_3$ atoms.

The band structure and spin expectation values, see Fig.~\ref{Fig:bands_spins_ABA_TMD}(b,c), are nicely fitted by our model Hamiltonian employing the parameters in Table~\ref{tab:fit}. 
The proximity-induced SOC parameters are also as expected, giving valley-Zeeman and opposite Rashba couplings for top and bottom graphene layers. 
Because the bottom graphene layer couples to MoSe$_2$, the values of the corresponding SOC parameters ($\lambda_{\textrm{I}}$ and $\lambda_{\textrm{R}}$) are smaller compared to the values for the top graphene layer, which couples to WSe$_2$. This result is in agreement with findings from single-layer-graphene/TMDC heterostructures \cite{Gmitra2016:PRB}. 
In addition, the middle layer experiences no proximity SOC, in agreement to BLG/TMDC results \cite{Zollner2020:PRL}, where only adjacent layers are proximitized. 
The two Rashba SOCs are opposite, because the bottom (top) graphene effectively feels the presence of a strong spin-orbit substrate (capping) layer, leading to opposite distortion of the corresponding $p_z$ orbitals.

\subsubsection{MoSe$_2$/ABC-TLG/WSe$_2$ stacks}

Next, we analyze the dispersion of the MoSe$_2$/ABC-TLG/WSe$_2$ heterostructure, see Fig.~\ref{Fig:bands_spins_ABC_TMD}(a). 
Again, the bands near the $K$ point closely resemble the ones from bare ABC-stacked TLG, but now with proximity SOC due to the surrounding TMDC layers. 
In contrast to ABA TLG, here we have 8 high-energy bands
at around $\pm 0.4$~eV, which are again formed by the dimer atoms in the structure, i. e., by the atoms 
A$_1$, A$_2$, B$_2$, and B$_3$ (see Fig.~\ref{Fig:pseudospin_ABC} in Appendix \ref{AppD}).
The remaining 4 low-energy bands are formed by non-dimer atoms B$_1$ and A$_3$, which are coupled by the hopping $\gamma_6$.
In addition, they feature a cone-like band touching along the $K\rightarrow M$ direction. 
Again, the bands and spin expectation values can be nicely reproduced by our model, see Fig.~\ref{Fig:bands_spins_ABC_TMD}(b,c), employing the fit parameters from Table~\ref{tab:fit}. 
Similar to BLG, the low-energy bands of ABC TLG can be strongly tuned by an electric field (in terms of band gap, see Fig.~\ref{Fig:bands_DOS_bare_ABC}) \cite{Konschuh2011:Diss}, since only two sublattice atoms from different layers contribute. The electric field tunability of the heterostructure dispersion will be investigated later. 
Similar to the ABA heterostructure, proximity-induced SOCs are of valley-Zeeman type and the outermost layers experience opposite Rashba couplings. 
The spin-orbit fields (see Fig.~\ref{Fig:spinfield_ABC}), corresponding to the low-energy bands $\varepsilon_{5-8}$, are mainly $s_z$ polarized, but show a Rashba texture especially around the cone-like touching points along $K\rightarrow M$ direction.
The band touching should actually determine the Fermi level. However in Fig.~\ref{Fig:bands_spins_ABC_TMD}, we have a small amount of doping ($E_D\neq 0$), due to the finite $k$-point sampling of the Brillouin zone in the DFT calculation. We therefore fix $E_D$ from the DFT, before we apply our fitting procedure.

\subsection{TLG encapsulated within magnetic Cr$_2$Ge$_2$Te$_6$}

\begin{figure*}[htb]
	\includegraphics[width=0.99\textwidth]{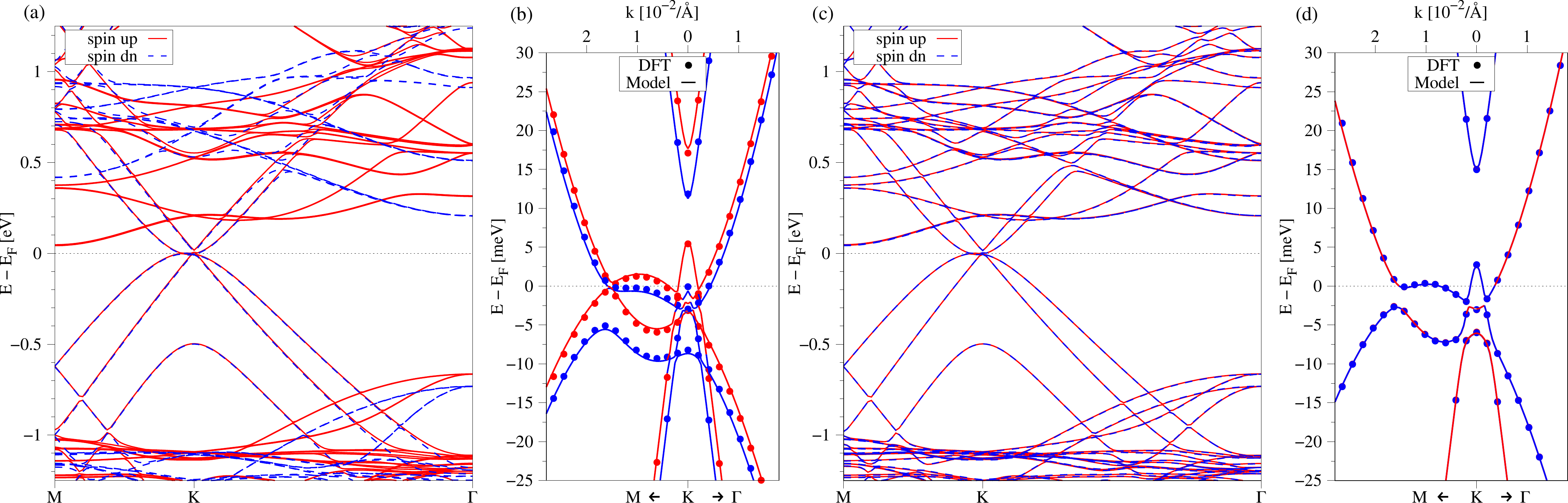}
	\caption{(a) Band structure of the CGT/ABA-TLG/CGT heterostructure along the $M-K-\Gamma$ path. Bands in solid red (dashed blue) correspond to spin up (spin down). The two CGT layers have parallel magnetizations along $z$ direction. 
 (b) Zoom to the calculated low-energy bands (symbols) around the $K$ point, corresponding to the band structure in (a), with a fit to the model Hamiltonian (solid lines). (c,d) Same as (a,b) but for antiparallel magnetization of the two CGT layers (bottom layer along $z$, top layer along $-z$).
 \label{Fig:bands_CGT_ABA}}
\end{figure*}

\begin{figure*}[htb]
	\includegraphics[width=0.99\textwidth]{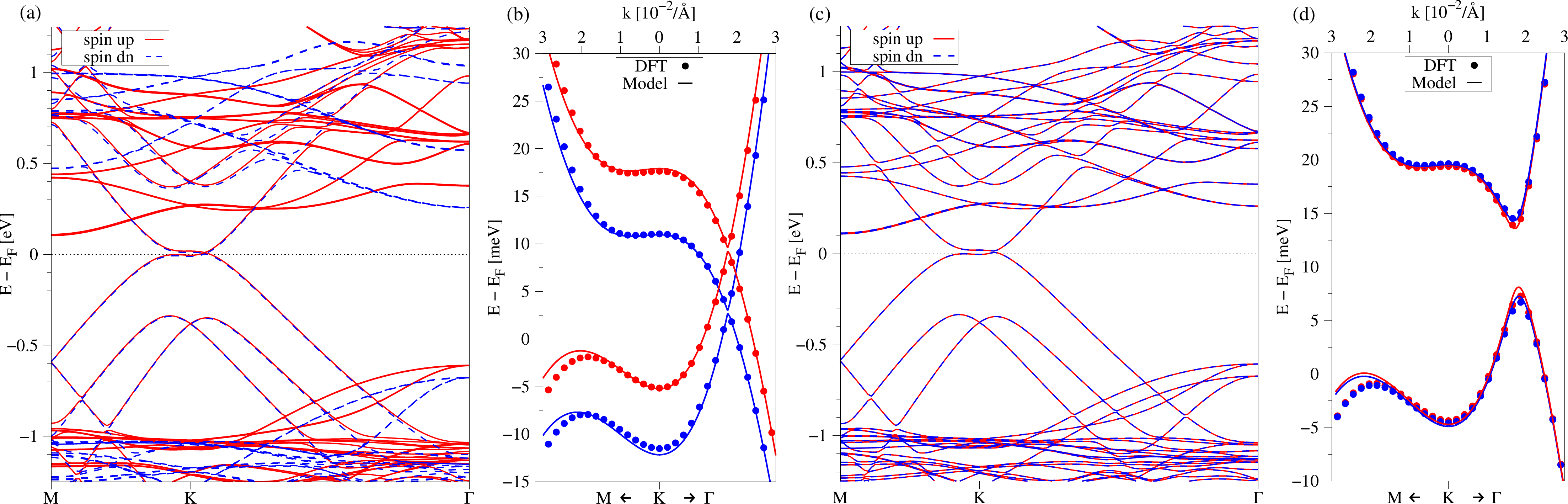}
	\caption{(a) Band structure of the CGT/ABC-TLG/CGT heterostructure along the $M-K-\Gamma$ path. Bands in solid red (dashed blue) correspond to spin up (spin down). The two CGT layers have parallel magnetizations along $z$ direction. 
 (b) Zoom to the calculated low-energy bands (symbols) around the $K$ point, corresponding to the band structure in (a), with a fit to the model Hamiltonian (solid lines). (c,d) Same as (a,b) but for antiparallel magnetization of the two CGT layers (bottom layer along $z$, top layer along $-z$).
 \label{Fig:bands_CGT_ABC}}
\end{figure*}

We now turn to proximity exchange, and consider TLG in ABA and ABC stacking encapsulated by the ferromagnetic semiconductor CGT. Again, the outermost graphene layers will experience proximity couplings, now stemming from the exchange interaction. 
In general, top and bottom graphene layers can feel different exchange fields. We investigate both parallel and antiparallel magnetizations of the encapsulating CGT layers and analyze the consequences for the TLG low-energy bands. 

\subsubsection{CGT/ABA/CGT stacks}

We first address the ABA TLG, sandwiched between two CGT layers. In Fig.~\ref{Fig:bands_CGT_ABA}(a) we show the calculated global band structure of the CGT/ABA-TLG/CGT heterostructure for parallel CGT magnetizations. Especially the ABA TLG low-energy bands can be nicely recognized near the Fermi level. The high-energy conduction bands of TLG are located within the CGT conduction bands. In addition, we find spin-polarized bands close to the Fermi level near the $M$ point, originating from the CGT layers. 
Zooming in on the Fermi level near the $K$ point, we again recognize the 8 low-energy bands originating from ABA TLG, see Fig.~\ref{Fig:bands_CGT_ABA}(b). Similar to the TMDC encapsulation, the bands are also significantly split, but now due to the exchange coupling originating from the ferromagnetic CGT layers. 
The bands can be nicely reproduced by our model Hamiltonian employing the fit parameters listed in Table~\ref{tab:fit}. In general, orbital parameters are barely affected by the surrounding materials proximitizing TLG, as can be seen by comparing with the fit results from the TMDC encapsultion. However, proximity-induced exchange couplings, $\lambda_{\textrm{ex}}$, are now necessary to capture the essential band structure features. From the fit, we find uniform exchange parameters of about $-3.4$~meV for all C sublattice atoms of the two outer graphene layers. This is in agreement with previous considerations for encapsulated BLG \cite{Zollner2021:arXiv,Zollner2020:PRL}.

What happens if we switch the magnetization direction of one CGT layer? In Fig.~\ref{Fig:bands_CGT_ABA}(c), we show the calculated global band structure of the CGT/ABA-TLG/CGT heterostructure for antiparallel CGT magnetizations. In general, the band structure remains the same, but bands originating from the topmost CGT layer have switched their spin-polarization. Most important are the consequences for the TLG low-energy bands, see Fig.~\ref{Fig:bands_CGT_ABA}(d). 
For parallel magnetizations, the bands were spin split due to uniform proximity-induced exchange couplings.
By switching the magnetization direction of the top CGT layer, the proximity exchange of the topmost graphene layer also adapts and switches sign, as reflected in the fitted parameters, see Table~\ref{tab:fit}.
Since the low-energy bands of TLG are equally formed by carbon atoms from top and bottom graphene layers, now with opposite proximity exchange couplings which effectively cancel each other, the bands remain spin degenerate. 
Again, this is similar as in CGT encapsulated BLG \cite{Zollner2021:arXiv}.

\subsubsection{CGT/ABC/CGT stacks}

Now, we turn to the CGT encapsulated ABC heterostructure. The band structure and fit results are summarized in Fig.~\ref{Fig:bands_CGT_ABC}, again considering parallel or antiparallel magnetizations of the CGT layers. The low-energy bands of ABC TLG are located near the Fermi level, similar to the TMDC encapsulation. The touching point of the bands is now along the $K\rightarrow \Gamma$ direction, see Fig.~\ref{Fig:bands_CGT_ABC}(b). In contrast, for the TMDC-encapsulated case, the touching was along the $K\rightarrow M$ direction, see Fig.~\ref{Fig:bands_spins_ABC_TMD}(b). This is related to the different supercell sizes of ABC TLG in the different encapsulation scenarios, and the backfolding of the graphene $k$ points into the heterostructure Brillouin zone \cite{Zhou2013:N}. 
A similar observation can be made considering the ABA TLG band structures for the different encapsulations. Actually, the touching point should determine the Fermi level, but due to the finite $k$-point sampling in our DFT calculation a small doping of about 5~meV appears. 

Most important is the low-energy band structure, featuring 4 Dirac-like bands, see Fig.~\ref{Fig:bands_CGT_ABC}(b). In the parallel magnetization case, the bands are split due to uniform proximity-induced exchange coupling in the outermost graphene layers of ABC TLG. The model parameters from Table~\ref{tab:fit} perfectly reproduce the dispersion in the vicinity of the $K$ point, featuring exchange parameters of about $-3.2$~meV. Looking at the fitted orbital parameters, we find that $\gamma_3$ and $\gamma_6$ have the opposite sign, but are nearly equal in value, compared to the TMDC-encapsulated ABC structure. This is again related to the different supercells and the backfolding, as already explained. 

What happens when we switch the magnetization direction of one CGT layer?
In the antiparallel case, see Fig.~\ref{Fig:bands_CGT_ABC}(c,d), the overall band structure is similar to the parallel magnetization. Again, bands originating from the topmost CGT layer have switched their spin-polarization. Most important are the consequences for the ABC TLG low-energy bands. 
They remain nearly spin degenerate, since the proximity exchange couplings from top and bottom graphene layer are again opposite in sign but nearly equal in value. In addition, a sizable gap opens at the former touching point, which originates from the layered antiferromagnetic proximity exchange in the outermost graphene layers of ABC TLG. 
This is similar to the observations we made in CGT encapsulated BLG \cite{Zollner2021:arXiv}, where cancellation of proximity exchange or proximity SOC leads to a gap opening without an external electric field. 
Therefore, also in the TMDC encapsulated scenario, we could open the gap if proximity SOCs from top and bottom graphene layers are opposite.
This happens, for example, when employing the same TMDC layers for the encapsulation of ABC TLG, but with a relative twist angle of $60^{\circ}$ between them. 

Given the similarity of the low energy physics of BLG and ABC TLG, the recently proposed spin valve device concepts \cite{Zhai2021:PRA,Wu2021:NJP,Cardoso2018:PRL}, which employ proximitized BLG, should be also applicable to ABC TLG. 
Most important, since the low energy band splittings depend on the magnetic configuration of the outer magnetic layers, as well as on the applied transverse electric field, one can efficiently control the in-plane conductance in doubly proximitized ABC-TLG.

\begin{table*}[htb]
\caption{\label{tab:fit} The fit parameters of the model Hamiltonian
$\mathcal{H}$ for the TMDC and the CGT encapsulated TLG structures. The arrows indicate the magnetization direction of CGT. }
\begin{ruledtabular}
\begin{tabular}{l c c c c c c }
\multirow{3}{*}{system} & WSe$_2$  & WSe$_2$  & CGT $\uparrow$  & CGT $\downarrow$ & CGT $\uparrow$  & CGT $\downarrow$ \\
 & ABC & ABA & ABC & ABC & ABA & ABA \\
 & MoSe$_2$  & MoSe$_2$  & CGT $\uparrow$  & CGT $\uparrow$ & CGT $\uparrow$  & CGT $\uparrow$ \\
\hline 
$\gamma_0$ [eV] &  2.5307 & 2.5482 & 2.5048 & 2.5200 & 2.5412 & 2.5375 \\
$\gamma_1$ [eV] &  0.4014 & 0.4028 & 0.3899 & 0.3898 & 0.3612 & 0.3585 \\
$\gamma_2$ [eV] &  0 & -0.0119 & 0 & 0 & -0.0099 & -0.0101 \\
$\gamma_3$ [eV] &  0.3292 & 0.3102 & -0.3287 & -0.3341 & -0.2926 & -0.2926 \\
$\gamma_4$ [eV] & -0.1838 &  -0.1765 & -0.1755 & -0.1778 & -0.1723 & -0.1727 \\
$\gamma_5$ [eV] & 0 & 0.0181 & 0 & 0 & 0.0076 & 0.0049 \\
$\gamma_6$ [eV] & 0.0125 & 0 & -0.0113 & -0.0116 & 0 & 0 \\
$V_1$ [meV] &  0.027 & 0.077 & 0.116 & -0.170 &  -0.530 & -0.076 \\
$V_2$ [meV] &  -11.022 & -6.278 & -12.045 & -7.469 & -3.264 & -2.952 \\
$\Delta$ [meV] & 7.961 & 22.084 & 11.095 & 6.937 & 10.162 & 7.583 \\
$\eta$ [meV] &  -3.121 & 6.572 & -2.831 & -3.076 & 4.494 & 4.484 \\
$\lambda_{\textrm{R1}}$ [meV] &  0.233 & 0.215 & 0 & 0 & 0 & 0\\
$\lambda_{\textrm{R3}}$ [meV] &  -0.475 & -0.459 & 0 & 0 & 0 & 0\\
$\lambda_{\textrm{I}}^\textrm{A1}$ [meV] &  0.209 & 0.204 & 0 & 0 & 0 & 0\\
$\lambda_{\textrm{I}}^\textrm{B1}$ [meV] &  -0.204 & -0.206 & 0 & 0 & 0 & 0 \\
$\lambda_{\textrm{I}}^\textrm{A3}$ [meV] &  1.125 &  0.995 & 0 & 0 & 0 & 0\\
$\lambda_{\textrm{I}}^\textrm{B3}$ [meV] &  -0.983 & -1.011 & 0 & 0 & 0 & 0\\
$\lambda_{\textrm{ex}}^\textrm{A1}$ [meV] &  0 & 0 & -3.393 & -3.393 & -3.210 & -3.210\\
$\lambda_{\textrm{ex}}^\textrm{B1}$ [meV] &  0 & 0 & -3.393 & -3.393 & -3.210 & -3.210\\
$\lambda_{\textrm{ex}}^\textrm{A3}$ [meV] &  0 & 0 & -3.349 & 3.349 & -3.210 & 3.210\\
$\lambda_{\textrm{ex}}^\textrm{B3}$ [meV] &  0 & 0 & -3.349 & 3.349 & -3.210 & 3.210\\
$E_D$ [meV] &  -4.171 & 0 & 5.540 & 10.605 & 0 & 0  \\
dipole [debye] & -0.0152 & -0.0109 & -0.0110 & 0.0010 & 0.0087 & 0.0101 \\
\end{tabular}
\end{ruledtabular}
\end{table*}

\section{Electric Field Tunability}
\label{efield}

Finally, we consider the electric field tunability of the low-energy bands of encapsulated TLG. The main effect of the field is that we place the outermost graphene layers in different potentials, thereby opening band gaps in the spectrum; see for example Fig.~\ref{Fig:bands_DOS_bare_ABC} for bare ABC TLG.
In the following, we restrict ourselves to the case of TMDC encapsulation, since for the CGT case one can expect similar outcomes. More precisely, as we have seen above, the most striking difference between the CGT and TMDC encapsulation cases is that proximity exchange coupling, rather than proximity SOC, is responsible for the band splittings. From the model Hamiltonian perspective, see Table~\ref{tab:fit}, orbital parameters are barely different for the two encapsulation scenarios, while proximity exchange couplings, $\lambda_{\textrm{ex}}$, are about 3-times larger in magnitude compared to proximity SOCs, $\lambda_{\textrm{I}}$. Therefore, one can expect a similar electric field behavior of the TLG low energy bands in both encapsulations. In addition, below we restrict ourselves to consider small field values of $\pm$0.25~V/nm, to capture the main effects of the field on the dispersion. In Fig.~\ref{Fig:Structure}, we specify the direction of a positive electric field.

\begin{figure}[htb]
	\includegraphics[width=0.99\columnwidth]{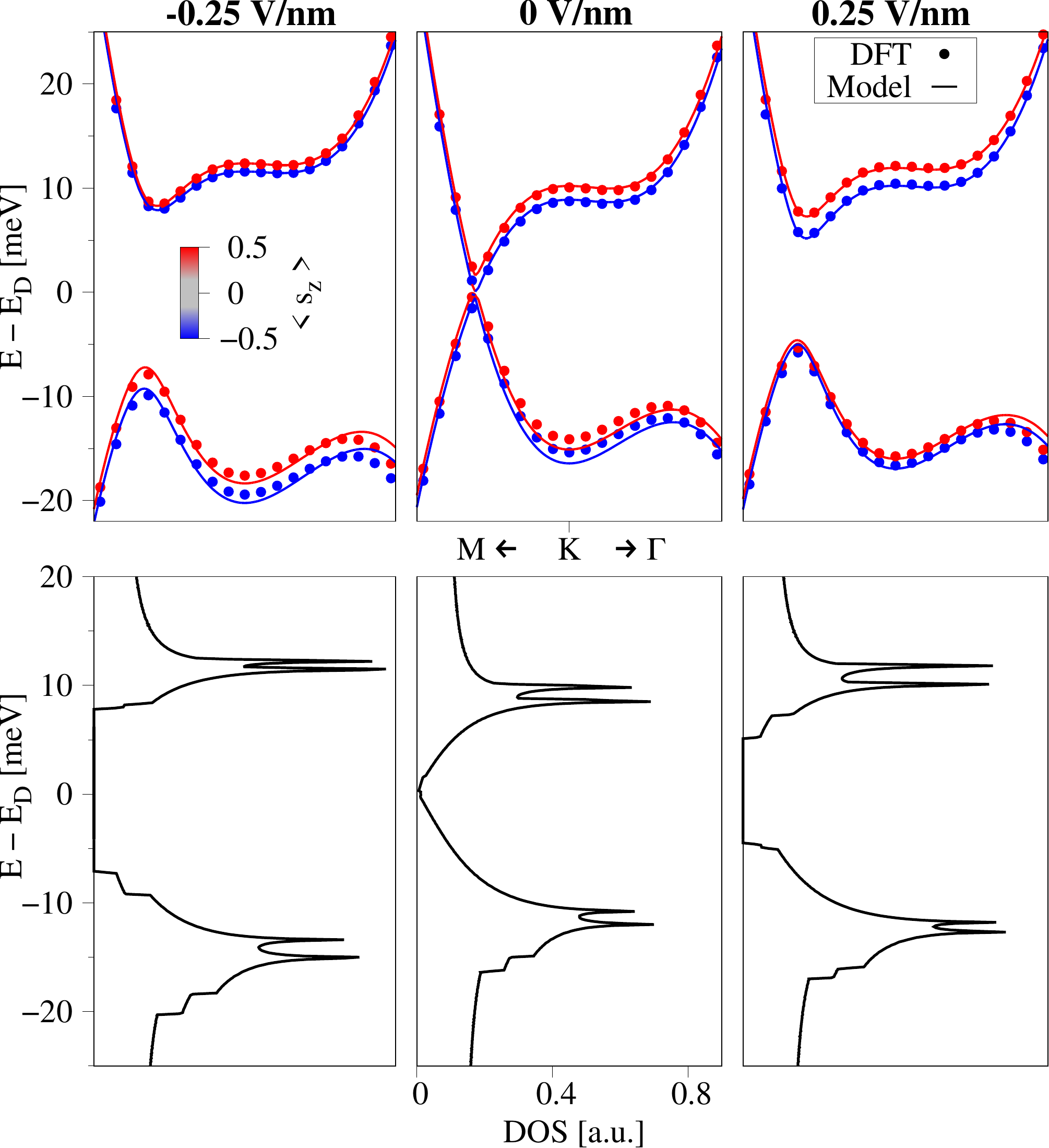}
	\caption{Top: Electric field evolution of the low-energy bands of the MoSe$_2$/ABC-TLG/WSe$_2$ heterostructure near the $K$ point. The color of the bands corresponds to the $s_z$ spin expectation value. From left to right, we tune the electric field from $-0.25$~V/nm to 0.25~V/nm. The calculated low-energy bands (symbols) match well with the model Hamiltonian fits (solid lines). Bands are plotted with respect to the Dirac point energy $E_D$. For the $\pm 0.25$~V/nm cases, we fixed the parameters from Table \ref{tab:fit}, and refitted the potential $V_1$ and asymmetry $\eta$, since mainly these parameters are affected by a small electric field (see Table~\ref{tab:fit_bare_ABC}). We obtained $V_1 = 6.259~(-3.839)$~meV and $\eta = -6.782~(-0.771)$~meV for the negative (positive) field. Bottom: The corresponding electric field evolution of the DOS, as calculated from the model Hamiltonian.
 \label{Fig:bands_ABC_Efield}}
\end{figure}

\subsection{MoSe$_2$/ABC-TLG/WSe$_2$ stacks}

In Fig.~\ref{Fig:bands_ABC_Efield} we show the electric field evolution of the MoSe$_2$/ABC-TLG/WSe$_2$ low-energy bands, including model fits.  
Since our model Hamiltonian very accurately reproduces the low-energy dispersions, we also show the corresponding DOS, calculated from the model \cite{dos}. 
When an electric field is applied across ABC TLG, a band gap opens at the cone-like band touching along the $K\rightarrow M$ line. 
The field separates the outer graphene layers in energy by the potential $V_1$. In order to fit the low-energy bands, when an electric field is applied, we keep the parameters from Table~\ref{tab:fit} fixed, but refit the parameters $\eta$ and $V_1$. The reason is that the applied field is small ($\pm0.25$~V/nm) and especially these two orbital parameters are relevant for the atoms B$_1$ and A$_3$, forming the low-energy bands. With this approach, the band structures with electric field are almost perfectly reproduced by the model.

In addition, we note that for negative field, the conduction band is split much less than the valence band and vice versa for the positive field. This can be also seen by looking at the calculated DOS. 
Such a switching of band splittings reminds us of the BLG physics \cite{Gmitra2017:PRL,Zollner2020:PRL,Island2019:Nat,Zollner2021:arXiv}. 
As theoretically proposed \cite{Gmitra2017:PRL} and experimentally demonstrated \cite{Island2019:Nat}, a transverse electric field, applied across a BLG/TMDC heterostructure, can efficiently tune SOC of conduction electrons, making it a potential platform for a spin-transistor. 
For ABC TLG, the low-energy bands are formed by atoms B$_1$ and A$_3$, which experience different magnitudes of proximity SOC, as the corresponding spin-orbit parameters in Table~\ref{tab:fit} show. For positive electric field, the bottom graphene layer, which couples to MoSe$_2$, sits in a lower potential than the top graphene layer, which couples to WSe$_2$. Therefore, the valence band is mainly formed by atoms B$_1$, experiencing proximity SOC of about 0.2~meV, while the conduction band is formed by atoms A$_3$, having proximity SOC of about 1~meV. For negative electric field, the situation is reversed allowing to fully electrically swap between the two magnitudes of SOC in the low-energy bands of ABC TLG. 
Such a swapping between spin-orbit splittings allows to electrically control spin relaxation in proximitized ABC TLG. 

\subsection{MoSe$_2$/ABA-TLG/WSe$_2$ stacks}

\begin{figure}[htb]
	\includegraphics[width=0.99\columnwidth]{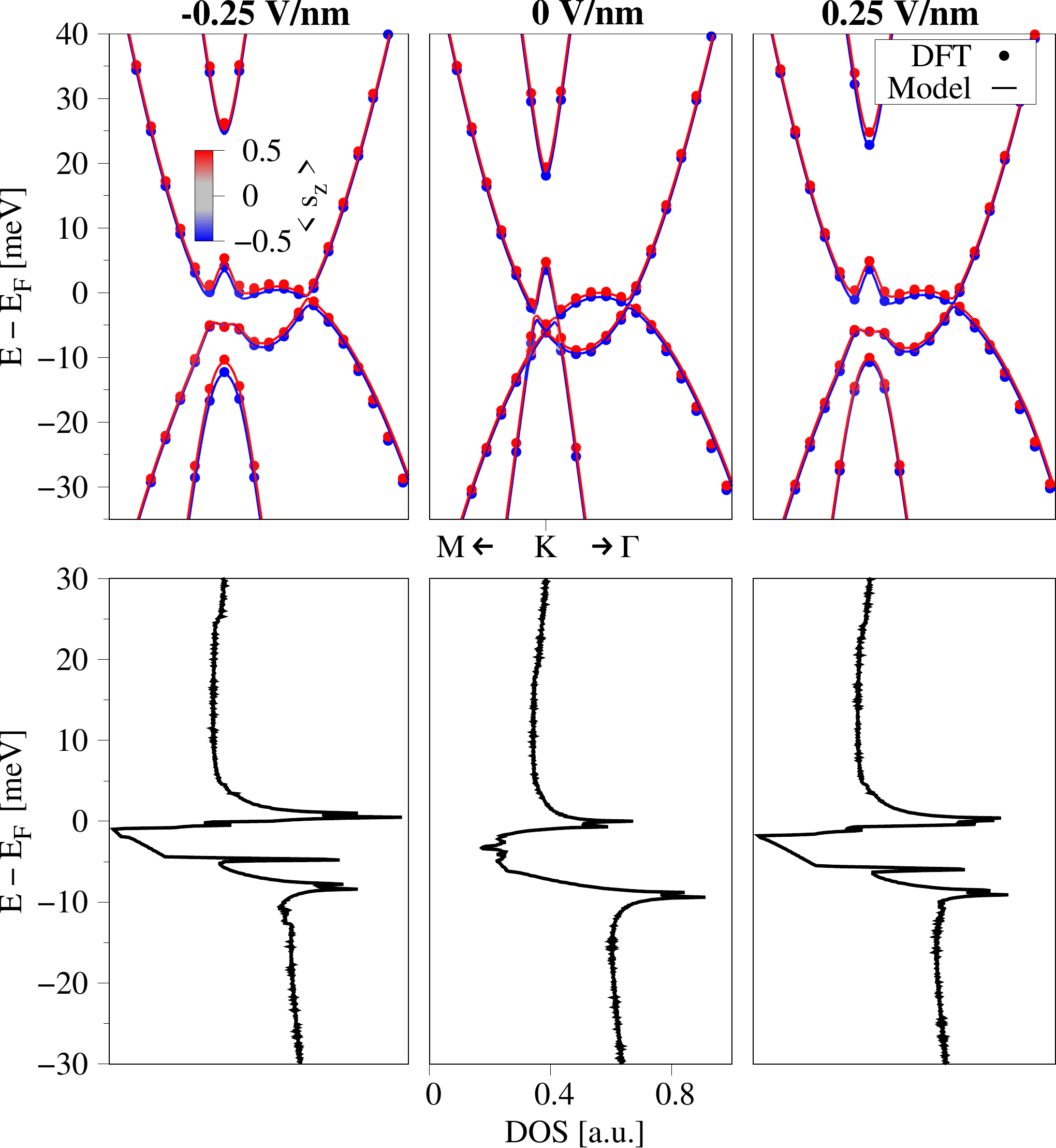}
	\caption{Top: Electric field evolution of the low-energy bands of the MoSe$_2$/ABA-TLG/WSe$_2$ heterostructure near the $K$ point. The color of the bands corresponds to the $s_z$ spin expectation value. From left to right, we tune the electric field from $-0.25$~V/nm to 0.25~V/nm. The calculated low-energy bands (symbols) match well with the model Hamiltonian fits (solid lines). For the $\pm 0.25$~V/nm cases, we fixed the parameters from Table \ref{tab:fit}, and refitted the potentials $V_1$ and $V_2$, since mainly these parameters are affected by a small electric field (see Table~\ref{tab:fit_bare_ABA}). We obtained $V_1 = 13.748~(-12.021)$~meV and $V_2 = -4.984~(-6.195)$~meV for the negative (positive) field. Bottom: The corresponding electric field evolution of the DOS, as calculated from the model Hamiltonian.
 \label{Fig:bands_ABA_Efield}}
\end{figure}

The impact of a transverse electric field on the ABA bands and DOS is more complicated than for the ABC case, see Fig.~\ref{Fig:bands_ABA_Efield}. 
Here, we have 8 low-energy bands that are present near the Fermi level, and which are formed by almost all carbon atoms from the three layers. Nevertheless, similar to ABC TLG, the electric field introduces a potential difference between the outermost graphene layers. In the model Hamiltonian, we capture this with parameter $V_1$. In addition, the middle layer can in general also sit in some nonzero potential, that we capture by $V_2$. Therefore, in order to fit the bands when an electric field is applied, we fix the parameters from Table~\ref{tab:fit}, but refit the parameters $V_1$ and $V_2$. Again, the band structures with electric field are also nicely reproduced by the model Hamiltonian, allowing us to calculate the DOS \cite{dos}.  

With the small field amplitudes that we employ, the low-energy dispersion barely changes. 
Nevertheless, the Dirac-like bands are further separated in energy and now strongly anticross with the BLG-like bands. The parabolic bands still touch along the $K\rightarrow \Gamma$ direction and the spectrum is not yet gapped for fields of $\pm 0.25$~V/nm, as can be seen from the DOS.

\section{Interplay of spin interactions}
\label{spininter}

One important conclusion of the previous discussion is that one can very well reproduce DFT data with our model Hamiltonians. The inclusion of an electric field, transverse to the monolayers, is straightforward, as it can be parametrized by essentially two orbital parameters. Considering also that the proximity effects are short ranged, we can realistically study more complex structures, such as ex-so-tic CGT/TLG/TMDC heterostructures \cite{exsotic}.

To be specific, we investigate TLG encapsulated by strong spin-orbit semiconductor WSe$_2$ on the top, and ferromagnetic semiconductor CGT on the bottom.  
To calculate the dispersion from the model Hamiltonian, we use the orbital parameters from Table~\ref{tab:fit} of TMDC-encapsulated ABA and ABC TLG. 
In addition, we employ valley Zeeman and Rashba SOC for the top graphene layer ($\lambda_{\textrm{I}}^\textrm{A3} = - \lambda_{\textrm{I}}^\textrm{B3} = 1$~meV, $\lambda_{\textrm{R3}} = -0.4$~meV), as well as exchange coupling for the bottom graphene layer ($\lambda_{\textrm{ex}}^\textrm{A1} =  \lambda_{\textrm{ex}}^\textrm{B1} = -3.5$~meV). 
This is a reasonable choice, according to our findings. 
To study the electric field behavior, also on the model level, we additionally employ the fitted parameters for the different electric fields ($V_1$ and $V_2$ for ABA; $V_1$ and $\eta$ for ABC), see Fig.~\ref{Fig:bands_ABC_Efield} and Fig.~\ref{Fig:bands_ABA_Efield}. 
The combination of SOC and proximity exchange breaks time-reversal symmetry. Therefore, we calculated the dispersion near both the $K$ and $K^{\prime}$ valleys. 

\subsection{ABC}
\begin{figure}[htb]
	\includegraphics[width=0.99\columnwidth]{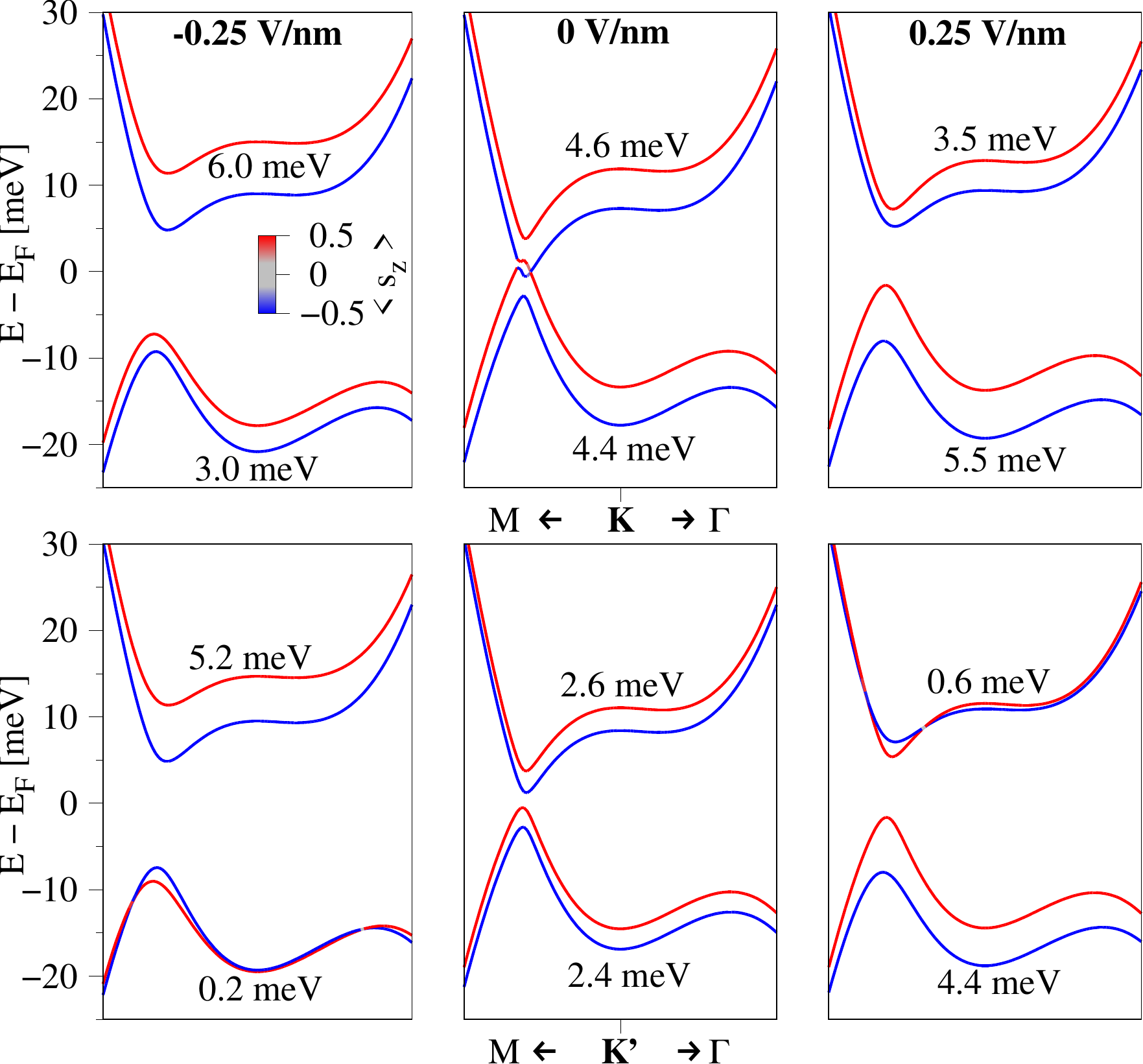}
	\caption{Electric field evolution of the low-energy bands of a CGT/ABC-TLG/WSe$_2$ heterostructure near the $K$ (top row) and $K^{\prime}$ (bottom row) points. The color of the bands corresponds to the $s_z$ spin expectation value. From left to right, we tune the electric field from $-0.25$~V/nm to 0.25~V/nm. 
	Next to the bands, we list the spin splitting at the $K$/$K^{\prime}$ point. The dispersions are calculated from the model, assuming reasonable parameters as explained in the text.
 \label{Fig:Exsotic_model_ABC}}
\end{figure}

The model results for ABC TLG are summarized in Fig.~\ref{Fig:Exsotic_model_ABC}.
For zero field, both sublattice atoms B$_1$ and A$_3$ are equally contributing to all four low-energy bands. The band splittings result now from the interplay of proximity spin-orbit and exchange couplings, originating from the individual layers. 
Consequently, at $K$ ($K^{\prime}$), the two spin interactions are additive (subtractive) and band splittings are about 4.5~meV (2.5~meV). 
Moreover, we find that the ABC low-energy bands exhibit no gap near $K$ valley, while there is a gap near $K^{\prime}$ valley, see Fig.~\ref{Fig:Exsotic_model_ABC}, potentially important for the realization of a valley-polarized quantum anomalous Hall effect \cite{Vila2021:arxiv}.

When an electric field is introduced, the bands are separated in energy, since the outermost layers are now in different potentials. In addition, the bands are not anymore equally formed by the two sublattice atoms. Still, the different spin interactions are at interplay, either in an additive or subtractive way. 
Consequently, for positive field both bands are spin split at $K$, while at $K^{\prime}$ the valence band is strongly split and the conduction band stays nearly degenerate. Reversing the direction of the electric field, the bands are flipped with respect to the Fermi level, since also the potential difference is now opposite compared to the positive field.
The fact that different magnitudes of splittings arise in different bands and valleys is important for the gate control of spin-relaxation \cite{Gmitra2017:PRL}, potentially resulting in valley- and spin-polarized currents \cite{Vila2021:arxiv}.

\subsection{ABA}

\begin{figure}[htb]
	\includegraphics[width=0.99\columnwidth]{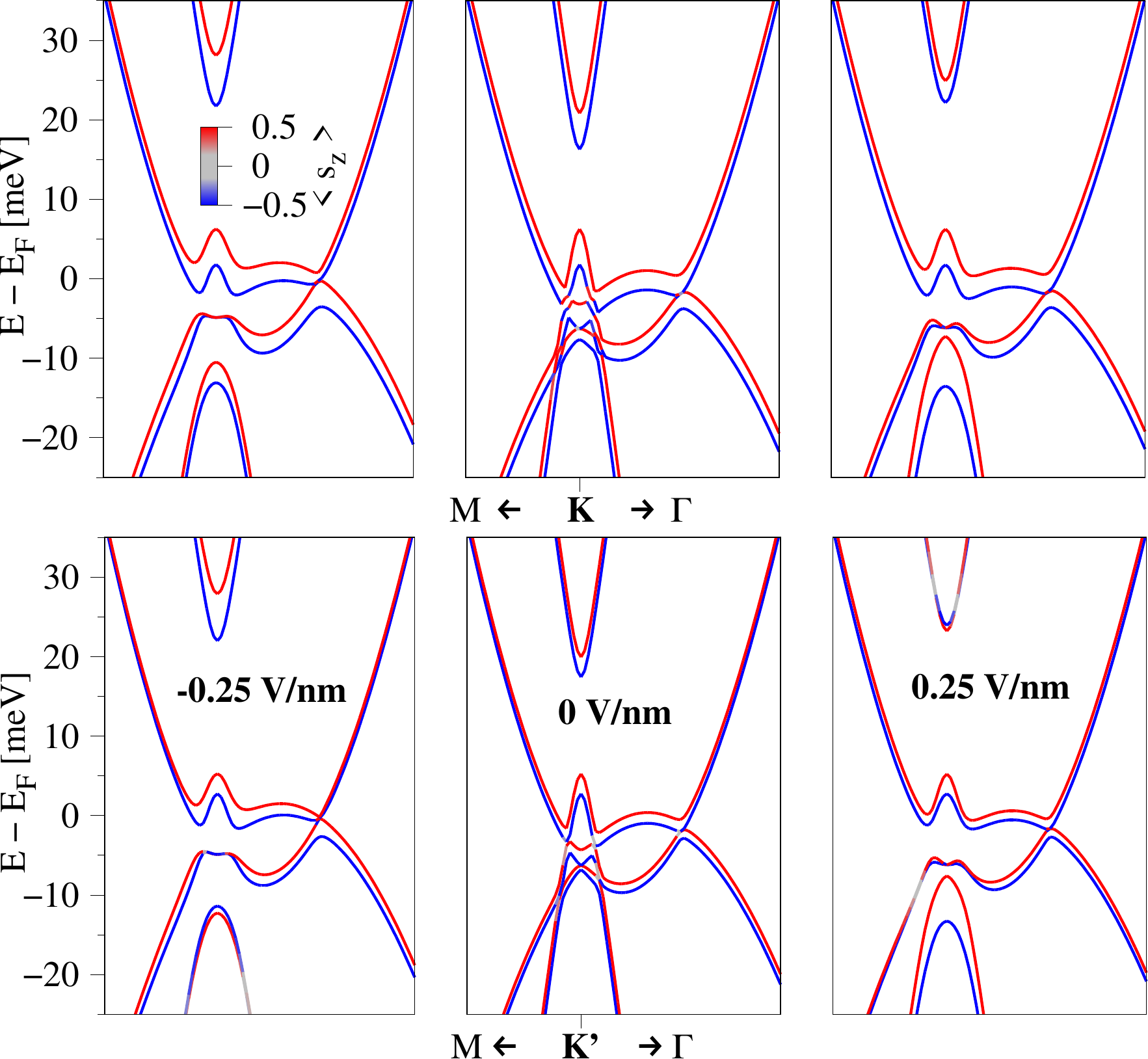}
	\caption{Electric field evolution of the low-energy bands of a CGT/ABA-TLG/WSe$_2$ heterostructure near the $K$ (top row) and $K^{\prime}$ (bottom row) points. The color of the bands corresponds to the $s_z$ spin expectation value. From left to right, we tune the electric field from $-0.25$~V/nm to 0.25~V/nm. The bands are calculated from the model, assuming reasonable parameters as explained in the text.
 \label{Fig:Exsotic_model_ABA}}
\end{figure}

In Fig.~\ref{Fig:Exsotic_model_ABA}, we show the calculated model low-energy dispersion of the ABA TLG, when the two spin interactions are at interplay. For zero field, we observe a similar behavior as for the ABC TLG, where near $K$ the band splittings are larger than near $K^{\prime}$, due to additive/subtractive effects of the spin interactions.  
In addition, looking at the field evolution ($-0.25 \rightarrow 0.25$~V/nm) of the Dirac-like bands, we observe a swapping of the nearly spin-degenerate bands from valence into conduction side at the $K^{\prime}$ valley, and vice versa for the strongly split bands. 
Similar observations can be made at the $K$ valley. The BLG-like parabolic bands remain more or less the same with the applied field. 

\section{Summary}
\label{summ}
We have investigated systematically the electronic band structure of TLG in different stacking and encapsulation configurations. Starting with bare ABA, ABC, and ABB TLG, we have provided the essential ingredients of the orbital and spin-orbital effects, and both quantitative and qualitative understanding of the SOC at the sub-meV level. Important for ongoing investigations, we also report on the electric-field tunability of the electronic DOS and VHS. 

The bulk of our investigation is devoted to encapsulated TLG. We deal separately with proximity SOC and proximity exchange. The proximity SOC is studied by encapsulating ABA and ABC TLG within WSe$_2$ and MoSe$_2$. We show how the two semiconducting monolayers affect the low-energy bands of TLG, inducing relatively strong (meV) spin splittings that are tunable by electric field. We also show that the effective model Hamiltonian, which considers sizeable SOC on the outer layers, performs great in explaining the DFT results. 

Proximity exchange is analyzed for the stacks of TLG encapsulated within CGT monolayers. The proximity exchange, also on meV scale, modifies the low-energy bands of TLG, differently for parallel and antiparallel CGT magnetizations. The magnetic tunability should be a useful way to control correlated phases, especially in ABC TLG which exhibits a large DOS at low energies. 

The nice comparison between the DFT results and the model makes us confident in applying the model itself in more complex situations, such as studying ex-so-tic CGT/TLG/TMDC heterostructures~\cite{exsotic}. The interplay of spin-orbit and exchange coupling leads to a different band structure at $K$ and $K^{\prime}$. In addition, the layer polarization then allows for a strong tunability of the bands with a transverse electric field, which is able to close or open spin splittings, or even switch the spin polarization, for a particular band in encapsulated ABC TLG.

In all our investigated DFT cases we provide fitting parameters to the effective model as a useful resource for subsequent investigations of transport and correlated physics based on realistic model simulations.


\acknowledgments
This work was funded by the Deutsche Forschungsgemeinschaft (DFG, German Research Foundation) SFB 1277 (Project No. 314695032), SPP 2244 (Project No. 443416183), and the European Union Horizon 2020 Research and Innovation Program under contract number 881603 (Graphene Flagship). M.~G. acknowledges financial support provided by Slovak Research and Development Agency provided under Contract No. APVV-20-0150 and by the Ministry of Education, Science, Research and Sport of the Slovak Republic provided under Grant No. VEGA 1/0105/20.

\appendix

\section{Electric field effects of bare ABA and ABC trilayer graphene}
\label{AppA}

\begin{table*}[htb]
\caption{\label{tab:fit_bare_ABA} The fit parameters of the orbital model Hamiltonian
$\mathcal{H}_{\textrm{orb}}^{\textrm{ABA}}$ for ABA TLG with electric field. The fitted band structure results are summarized in Fig.~\ref{Fig:bands_DOS_bare_ABA}. }
\begin{ruledtabular}
\begin{tabular}{l c c c c c }
el. field [V/nm] & 0 & 0.25 & 0.5 & 0.75 & 1.0 \\
\hline 
$\gamma_0$ [eV] & 2.5692 & 2.5693 & 2.5695 &  2.5693 & 2.5688 \\
$\gamma_1$ [eV] & 0.3699 & 0.3700 & 0.3700 & 0.3700 & 0.3701 \\
$\gamma_2$ [eV] & -0.0101 &  -0.0101 & -0.0100  & -0.0099 &  -0.0098\\
$\gamma_3$ [eV] &  0.2829 &  0.2821 &  0.2798 & 0.2772 & 0.2742 \\
$\gamma_4$ [eV] & -0.1643 & -0.1640 &  -0.1638 &  -0.1642 & -0.1635  \\
$\gamma_5$ [eV] & 0.0150 & 0.0130 & 0.0129 & 0.0130 & 0.0132 \\
$V_1$ [meV] & 0 & -19.146 & -38.608 & -57.927 & -78.221 \\
$V_2$ [meV] & -19.991 & -14.814 & -13.980 &  -13.846 &  -14.715\\
$\Delta$ [meV] &  20.566 & 24.193 & 24.452  & 24.158 & 23.983 \\
$\eta$ [meV] &  6.965 & 12.842 & 13.189 & 12.757 &  12.506 \\
dipole [debye] & 0 & 0.0250 & 0.0499 & 0.0749 & 0.0996 \\
\end{tabular}
\end{ruledtabular}
\end{table*}

\begin{table*}[htb]
\caption{\label{tab:fit_bare_ABC} The fit parameters of the orbital model Hamiltonian
$\mathcal{H}_{\textrm{orb}}^{\textrm{ABC}}$ for ABC TLG with electric field. The fitted band structure results are summarized in Fig.~\ref{Fig:bands_DOS_bare_ABC}.  }
\begin{ruledtabular}
\begin{tabular}{l c c c c c }
el. field [V/nm] & 0 & 0.25 & 0.5 & 0.75 & 1.0 \\
\hline 
$\gamma_0$ [eV] & 2.5539 &  2.5560 & 2.5656 & 2.5884 &  2.6198 \\
$\gamma_1$ [eV] & 0.3690 & 0.3689 & 0.3689 &  0.3687 & 0.3685 \\
$\gamma_3$ [eV] & 0.2917  &  0.2948 &  0.2960 &  0.2905  & 0.2816 \\
$\gamma_4$ [eV] & -0.1646 &  -0.1642 & -0.1618  &  -0.1571 & -0.1516\\
$\gamma_6$ [eV] & 0.0108 &  0.0113 &  0.0117 & 0.0115  & 0.0111 \\
$V_1$ [meV] & 0  & -3.050  &  -10.124 &  -18.172 & -25.409 \\
$V_2$ [meV] & -37.243 & -37.184  &  -36.066 &  -34.365 & -32.529 \\
$\Delta$ [meV] & 10.288 & 9.746  &  9.001 & 9.816  & 9.094 \\
$\eta$ [meV] &  -1.121 & -1.551  & -1.664  & 0.172  & 0.467 \\
dipole [debye] & 0 & 0.0286  &  0.0552 &  0.0812 & 0.1077 \\
\end{tabular}
\end{ruledtabular}
\end{table*}

    \begin{figure*}[htb]
     \includegraphics[width=.99\textwidth]{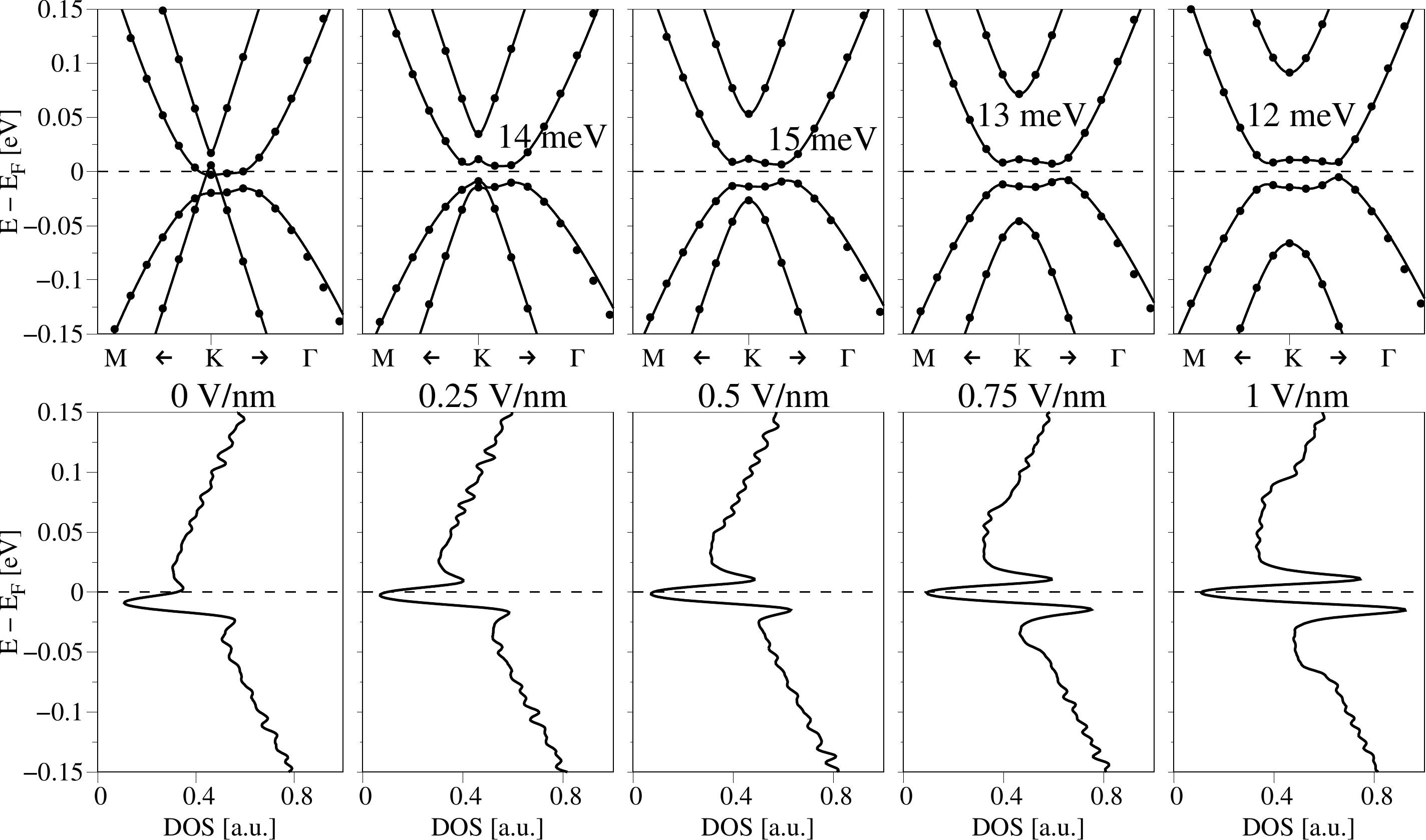}
     \caption{Top: Zooms to the ABA TLG bands in the vicinity of the $K$ point. We compare DFT data (symbols) with the model fits (solid lines) employing parameters from Table~\ref{tab:fit_bare_ABA}. In the dispersion, we list the band gap at the Fermi level. Bottom: The corresponding calculated density of states (DOS). From left to right, we increase the transverse electric field from 0 to 1~V/nm. 
     }\label{Fig:bands_DOS_bare_ABA}
    \end{figure*}

The following electric field results for bare ABA and ABC TLG have been obtained with \textsc{Quantum ESPRESSO} \cite{Giannozzi2009:JPCM}. 
We used the same structural input as for WIEN2k \cite{Wien2k}, as described in the main text. 
Self-consistent calculations are performed with the $k$-point sampling of 
$240\times 240\times 1$ to accurately determine the Fermi level and the DOS.
We used an energy cutoff for the charge density of $450$~Ry, and
the kinetic energy cutoff for wavefunctions is $55$~Ry for the scalar relativistic pseudopotentials 
with the projector augmented wave method \cite{Kresse1999:PRB} with the Perdew-Burke-Ernzerhof exchange correlation functional \cite{Perdew1996:PRL}. Moreover, we added
vdW \cite{Grimme2006:JCC,Barone2009:JCC} and
dipole corrections \cite{Bengtsson1999:PRB}.

Since we are mainly interested in the electric field behavior of the dispersion and the DOS, we neglect SOC, since the induced splittings are on the $\mu$eV level (see main text). In any case, the plane wave and pseudopotential method, implemented in \textsc{Quantum ESPRESSO} \cite{Giannozzi2009:JPCM}, cannot give correct spin-orbit splittings in graphene, since the relevant $d$-orbitals are missing \cite{Gmitra2009:PRB,Konschuh2010:PRB}. Nevertheless, on an orbital level, the dispersions are the same as calculated with WIEN2k \cite{Wien2k}, and we can safely study the electric field tunability. 

The fit parameters from Table~\ref{tab:fit_bare_ABA} and Table~\ref{tab:fit_bare_ABC}, nicely reproduce the DFT-calculated band structures in Fig.~\ref{Fig:bands_DOS_bare_ABC} for ABC TLG and Fig.~\ref{Fig:bands_DOS_bare_ABA} for ABA TLG, employing the orbital model Hamiltonians, $\mathcal{H}_{\textrm{orb}}^{\textrm{ABA}}$ and $\mathcal{H}_{\textrm{orb}}^{\textrm{ABC}}$, from the main text. 
The electric field behaviour of the ABC TLG dispersion and DOS was already discussed in the main text. 
For the ABA TLG, the electric field also introduces a potential difference between the outermost layers, separating the monolayer graphene bands in energy. The BLG bands stay more or less the same with the applied field, see Fig.~\ref{Fig:bands_DOS_bare_ABA}. In addition, a small band gap is introduced at the Fermi level. However, the gap first increases and then decreases again, when tuning the field from 0 to 1~V/nm. From the DOS, we find VHS associated with the parabolic BLG bands. With increasing field, we find that these bands even further flatten near the Fermi level, which comes along with increasing VHS in the DOS.

The fitted electric field results, summarized in Table~\ref{tab:fit_bare_ABA} and Table~\ref{tab:fit_bare_ABC} are valuable for further theoretical considerations, e.~g., gate-tunable transport simulations of bare ABA and ABC TLG.
If one wants to add SOC, this can be easily done on a model level, employing the intrinsic SOC of bare graphene, $\lambda_{\textrm{I}} = 12~\mu$eV, for all sublattice atoms \cite{Gmitra2009:PRB,Konschuh2010:PRB}. The Rashba SOC would be even smaller (at maximum 5~$\mu$eV) for our investigated field values \cite{Konschuh2011:Diss,Konschuh2012:PRB}. 
Certainly, one can also add proximity spin-orbit or exchange couplings, as we describe it for our encapsulated structures.

\section{Bare ABB trilayer graphene}
\label{AppB}
In addition to ABA and ABC TLG, there is also another stacking sequence, the ABB one. All those stackings are present in recent investigations on twisted-graphene/BLG heterostructures \cite{Polshyn2020:Nat,Rademaker2020:PRR,Park2020:PRB,Ma2021:SB}.
In the same manner as above and in the main text, we calculated the dispersion, DOS, and analyze the electric field behavior of ABB TLG.
A reasonable orbital Hamiltonian for ABB TLG can be constructed from combining the knowledge about AA and AB stacked BLG, together with the ABC and ABA TLG \cite{Konschuh2012:PRB,Konschuh2011:Diss,Rakhmanov2012:PRL,Rozhkov2016:PR}. We propose the following ABB TLG orbital Hamiltonian

\begin{widetext}
\begin{flalign}
\mathcal{H}_{\textrm{orb}}^{\textrm{ABB}} = & \begin{pmatrix}
\Delta+V_1 & \gamma_0 f(\bm{k}) & \gamma_4 f^{*}(\bm{k}) & \gamma_1 & 0 & \gamma_5 \\
\gamma_0 f^{*}(\bm{k}) & V_1 & \gamma_3 f(\bm{k}) & \gamma_4 f^{*}(\bm{k}) & 0 & 0 \\
 \gamma_4 f(\bm{k}) & \gamma_3 f^{*}(\bm{k}) & \Delta+V_2 & \gamma_0 f(\bm{k}) & \gamma_1^{\prime} & \gamma_7 f(\bm{k}) \\
\gamma_1 & \gamma_4 f(\bm{k}) & \gamma_0 f^{*}(\bm{k}) & \Delta+V_2 & \gamma_7 f(\bm{k}) & \gamma_1^{\prime} \\
0 & 0 & \gamma_1^{\prime} & \gamma_7 f^{*}(\bm{k}) &  \Delta-V_1 & \gamma_0 f(\bm{k}) \\
\gamma_5 & 0 & \gamma_7 f^{*}(\bm{k}) & \gamma_1^{\prime} & \gamma_0 f^{*}(\bm{k}) & \Delta-V_1
\end{pmatrix} \otimes s_0.
\end{flalign}
\end{widetext}

\begin{figure}[htb]
	\includegraphics[width=0.65\columnwidth]{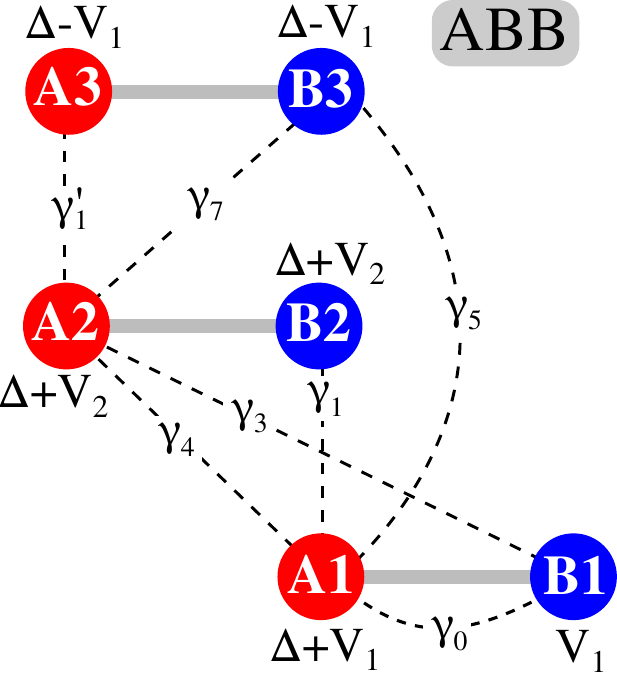}
	\caption{Schematic illustration of the ABB TLG lattice, showing the relevant intra- and interlayer hoppings $\gamma_j$, $j = \{ 0,1,3,4,5,7 \}$ (dashed lines). In addition, the bottom (top) graphene layer is placed in the potential $V_1$ ($-V_1$), while the middle layer is placed in potential $V_2$. The asymmetry $\Delta$ arise due to vertical hoppings. In general, the direct interlayer couplings $\gamma_1$ and $\gamma_1^{\prime}$ can be different. However, when both interlayer distances are the same (in our case fixed to $d = 3.3$~\AA), $\gamma_1 = \gamma_1^{\prime}$ holds. 
 \label{Fig:ABB_scheme}}
\end{figure}

\begin{figure}[htb]
	\includegraphics[width=0.98\columnwidth]{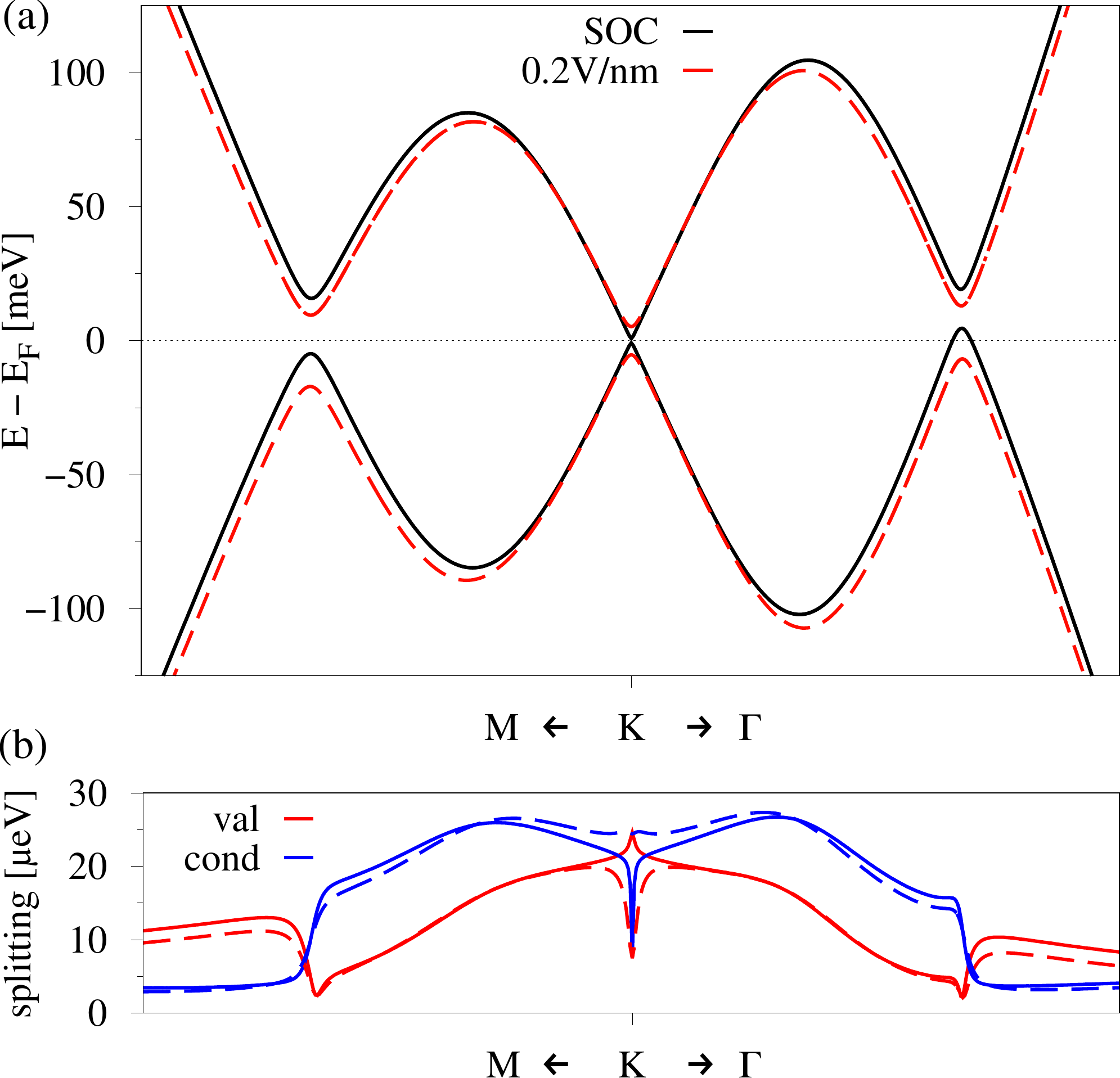}
	\caption{(a) DFT-calculated low-energy bands of ABB TLG with SOC (black solid line) and additionally with a transverse electric field of 0.2~V/nm (red dashed line). 
	(b) The spin splitting of valence (red) and conduction (blue) bands. Solid lines are with SOC only and dahed lines are with SOC in the presence of the electric field. 
 \label{Fig:Efield_ABB}}
\end{figure}

\begin{table*}[htb]
\caption{\label{tab:fit_bare_ABB} The fit parameters of the model Hamiltonian
$\mathcal{H}_{\textrm{orb}}^{\textrm{ABB}}$ for ABB TLG with electric field, assuming $\gamma_1 = \gamma_1^{\prime}$. The fitted band structure results are summarized in Fig.~\ref{Fig:bands_DOS_bare_ABB}.  }
\begin{ruledtabular}
\begin{tabular}{l c c c c c }
el. field [V/nm] & 0 & 0.25 & 0.5 & 0.75 & 1.0 \\
\hline 
$\gamma_0$ [eV] & 2.5662 &  2.5682 & 2.5713 & 2.5737 &  2.5754 \\
$\gamma_1$ [eV] & 0.3698 & 0.3697 & 0.3693 &  0.3687 & 0.3683 \\
$\gamma_3$ [eV] & 0.2726  &  0.2723 &  0.2716 &  0.2706  & 0.2694 \\
$\gamma_4$ [eV] & -0.1716 &  -0.1699 & -0.1722  &  -0.1760 & -0.1793\\
$\gamma_5$ [eV] & 0.0229 &  0.0228 &  0.0246 & 0.0264  & 0.0275 \\
$\gamma_7$ [eV] & -0.1087 &  -0.1101 &  -0.1053 & -0.0966  & -0.0874 \\
$V_1$ [meV] & 0  & -12.982  &  -31.824 &  -48.449 & -64.292 \\
$V_2$ [meV] & -51.533 & -53.755  &  -60.330 &  -66.823 & -71.386 \\
$\Delta$ [meV] & 21.367 & 21.737  &  20.917 & 19.660 & 18.542 \\
dipole [debye] & 0 & 0.0245  &  0.0500 &  0.0764 & 0.1029 \\
\end{tabular}
\end{ruledtabular}
\end{table*}

    \begin{figure*}[htb]
     \includegraphics[width=.99\textwidth]{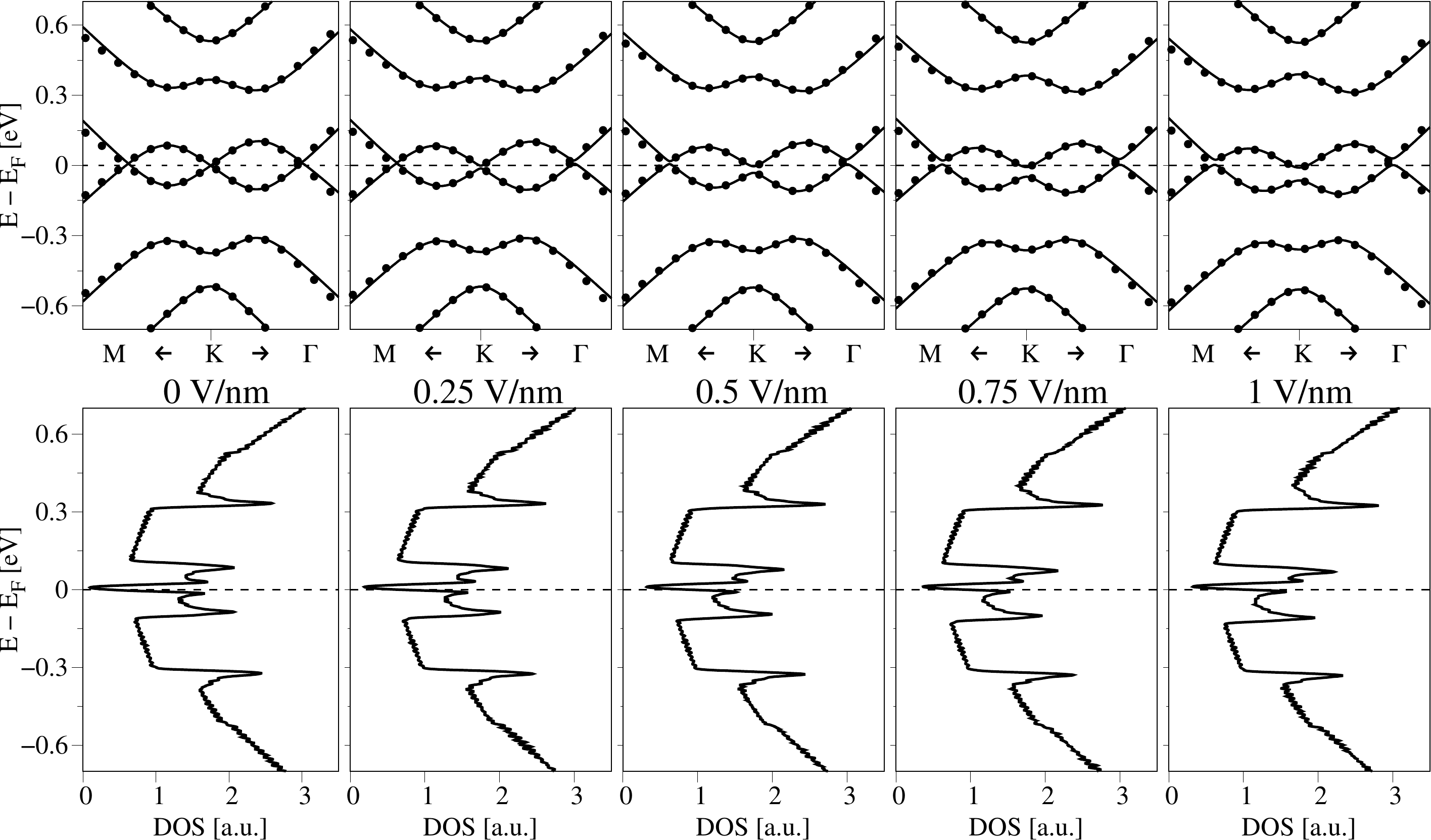}
     \caption{Top: Zooms to the ABB TLG bands in the vicinity of the $K$ point. We compare DFT data (symbols) with the model fits (solid lines) employing parameters from Table~\ref{tab:fit_bare_ABB}. Bottom: The corresponding calculated density of states (DOS). From left to right, we increase the transverse electric field from 0 to 1~V/nm. 
     }\label{Fig:bands_DOS_bare_ABB}
    \end{figure*}

A schematic illustration of the couplings in ABB TLG is shown in Fig.~\ref{Fig:ABB_scheme}.
The hoppings between the bottom and middle graphene layer are the same as for the other TLG structures. 
However, the hoppings between the middle and the top graphene layer are adapted from the AA BLG structure \cite{Rozhkov2016:PR}.
Since the ABB TLG has no $z$-mirror or inversion symmetry, connecting the top and bottom layer, the direct interlayer hoppings $\gamma_1$ and $\gamma_1^{\prime}$ can be in principle different. Only when the interlayer distances are the same, also the hopping amplitudes are the same.
The coupling $\gamma_7$, connecting opposite sublattices from different layers, is similar to the coupling $\gamma_4$, connecting same sublattices from different layers. Nevertheless, they can be very different in amplitude.

The pseudospin character and the spin-orbit splittings in the bare ABB TLG dispersion have been already discussed in the main text. 
In Fig.~\ref{Fig:Efield_ABB}, we show a zoom to the ABB TLG low-energy bands, further emphasizing the effect of a transverse electric field on the dispersion and the spin splittings. 
Applying the field, apparently flips the low-energy band splittings at the $K$ point with respect to the Fermi level. 
This is not surprising, considering the projected band structure in Fig.~\ref{Fig:bare_ABB}(c). Without the field, the low-energy conduction (valence) band has a pronounced contribution from B$_1$ (B$_3$) atoms near the anticrossing points away from $K$. With the electric field, applied along positive $z$ direction, a potential difference between the outermost layers is introduced, such that the bottom layer is now in the lowest potential. Therefore, the band characters of valence and conduction bands flip, along with the spin splittings near the $K$ point.

In Fig.~\ref{Fig:bands_DOS_bare_ABB}, we summarize the electric field behavior of the band structure and DOS for the ABB TLG. The dispersion can be nicely reproduced by the orbital model Hamiltonian, employing the fit results from Table~\ref{tab:fit_bare_ABB}.

\section{Effects of relaxation on bare TLG dispersions}
\label{AppC}

In the main text, when discussing bare TLG, we have fixed the interlayer distances between the graphene layers to $d = 3.3$~\AA, which is an approximation. 
We now allow the TLG structures to minimize their energy by relaxing the C atom positions. For the relaxation we employ DFT-D2 vdW corrections \cite{Grimme2006:JCC,Barone2009:JCC} and use quasi-newton algorithm based on trust radius procedure. To determine the interlayer distances, the carbon atoms are allowed to relax only in their $z$ positions (vertical to the layers), until all components of all forces are reduced below $10^{-4}$~[Ry/$a_0$], where $a_0$ is the Bohr radius. 

\begin{figure}[htb]
     \includegraphics[width=.99\columnwidth]{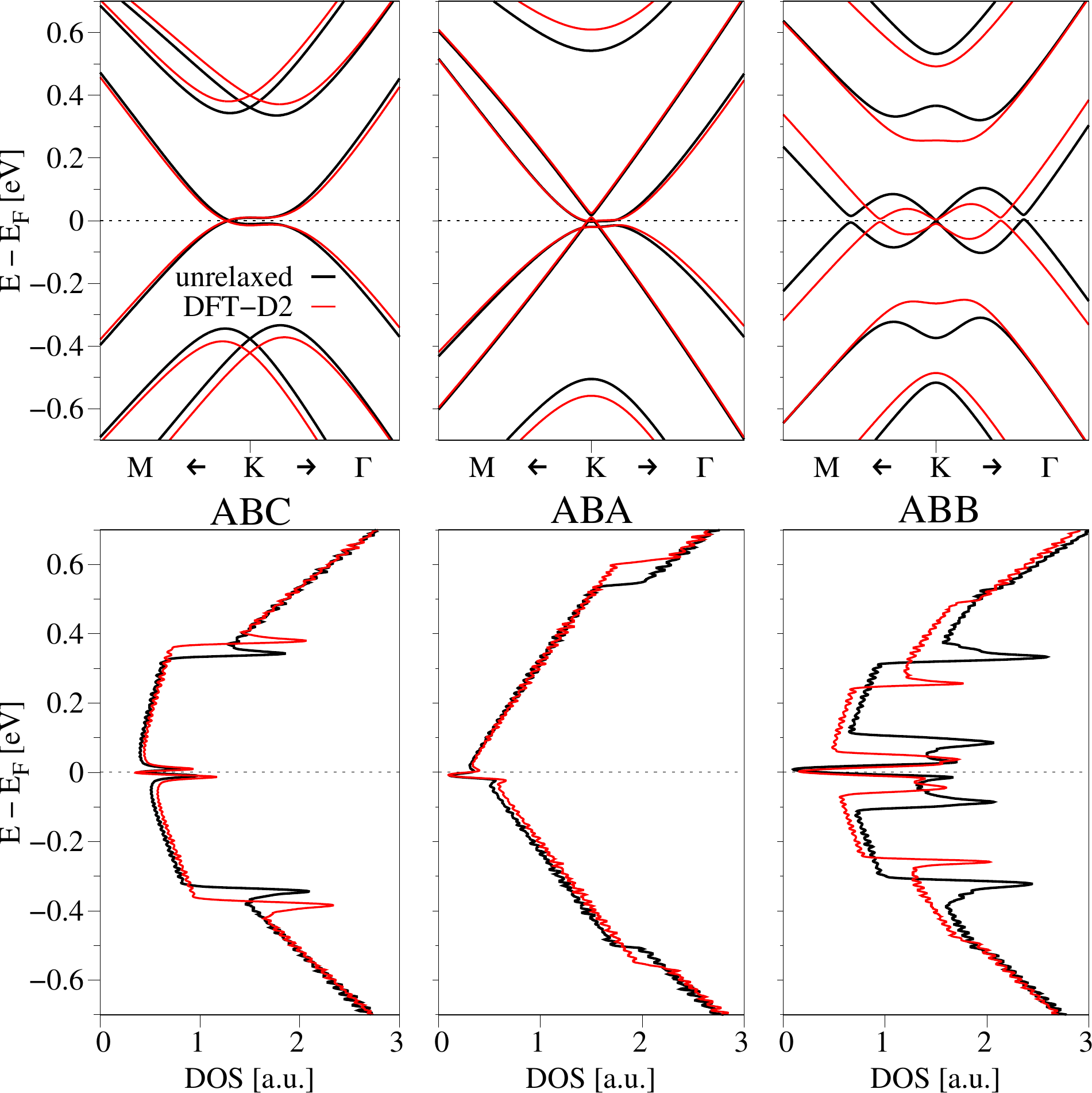}
     \caption{Top: Zooms to the TLG dispersions in the vicinity of the $K$ point. We compare unrelaxed (black) and fully relaxed (red) structures employing DFT-D2 vdW corrections. Bottom: The corresponding calculated density of states (DOS). The red lines can be reproduced (not shown) by model Hamiltonian parameters summarized in Table~\ref{tab:fit_relaxed_TLG}.
     }\label{Fig:bands_DOS_relaxed}
    \end{figure}

After relaxation, interlayer distances are slightly reduced to about $d = 3.24$~\AA. The only exception is the interlayer distance between BB-layers in ABB TLG, which is relaxed to about $d = 3.50$~\AA.
With the relaxed interlayer distances it is now also reasonable to compare the total energies of all the stackings. We find that the ABA TLG is energetically most favorable. However, the ABC one is only about $60~\mu$eV higher in energy. 
The ABB TLG is 25~meV higher in energy than the ABA one, making this stacking less favorable to be observed naturally. 
In Fig.~\ref{Fig:bands_DOS_relaxed}, we compare the band structure and DOS results for the TLG structures with and without the relaxation. The most drastic changes can be seen in the ABB TLG dispersion, since interlayer distances are now highly asymmetric. For the ABC and ABA TLG structures, the dispersion is barely different, except for the high-energy bands. 

\begin{table}[htb]
\caption{\label{tab:fit_relaxed_TLG} The fit parameters of the model Hamiltonians
for the relaxed TLG structures. The dispersions are shown in Fig.~\ref{Fig:bands_DOS_relaxed}.}
\begin{ruledtabular}
\begin{tabular}{l c c c }
TLG & ABC & ABA & ABB \\
\hline 
$\gamma_0$ [eV] & 2.5470 &  2.5687 & 2.5845  \\
$\gamma_1$ [eV] & 0.4103 & 0.4128 & 0.4133  \\
$\gamma_1^{\prime}$ [eV] & - & - & 0.2593  \\
$\gamma_2$ [eV] & - & -0.0067 & -  \\
$\gamma_3$ [eV] & 0.3243  &  0.3128 &  0.2755  \\
$\gamma_4$ [eV] & -0.1794 &  -0.1954 & -0.2018  \\
$\gamma_5$ [eV] & - &  0.0089 &  0.0167  \\
$\gamma_6$ [eV] & 0.0133 &  - &  -  \\
$\gamma_7$ [eV] & - &  - &  -0.0077  \\
$V_1$ [meV] & 0  & 0 &  0 \\
$V_2$ [meV] & -45.161 & -20.831  &  -27.767  \\
$\Delta$ [meV] & 11.105 & 31.501  &  9.040  \\
$\eta$ [meV] & -2.838 & 5.069 & - \\
\end{tabular}
\end{ruledtabular}
\end{table}

In Table~\ref{tab:fit_relaxed_TLG}, we summarize the fit results, when the structures are relaxed. Especially the direct interlayer couplings $\gamma_1$ and $\gamma_1^{\prime}$ are strongly renormalized, because of the different interlayer distances. The high-energy bands are split off exactly due to these couplings, explaining the differences in the dispersions with and without relaxation.

\section{Pseudospin character of TMDC encapsulated TLG dispersions}
\label{AppD}

\begin{figure*}[htb]
     \includegraphics[width=.8\textwidth]{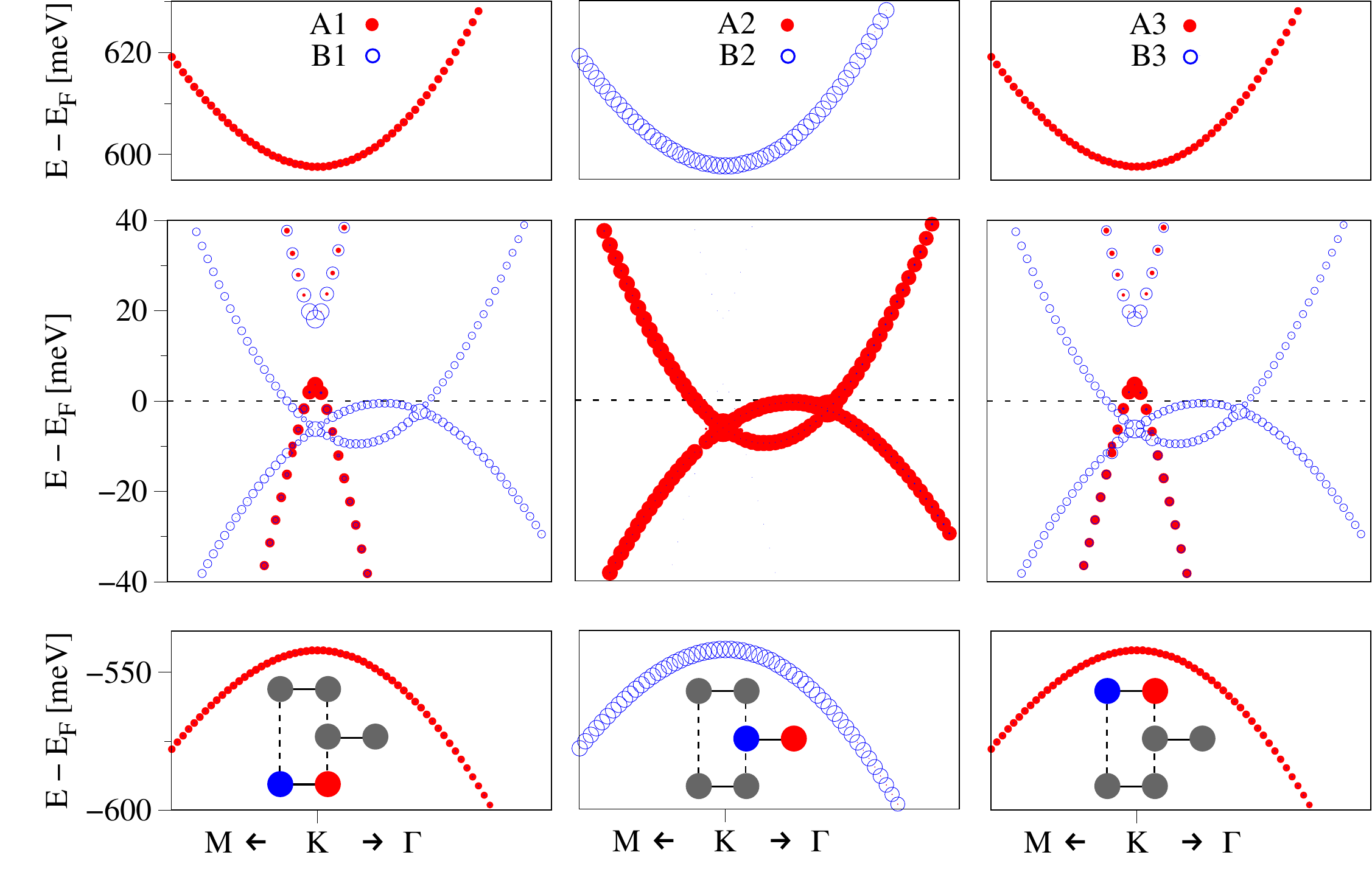}
     \caption{Zooms to the TLG bands in the vicinity of the $K$ point of the MoSe$_2$/ABA-TLG/WSe$_2$ heterostructure. From left to right, we project onto the sublattice atoms (A = red-filled circles, B = blue-open circles) of the bottom, middle, and top graphene layers. The sublattice atoms are indicated in the sketch of the ABA TLG geometry in the lower panels.
     }\label{Fig:pseudospin_ABA}
    \end{figure*}

In Fig.~\ref{Fig:pseudospin_ABA}, we show the pseudospin character of the relevant ABA TLG bands of the TMDC encapsulated heterostructure. Similar as in Fig.~\ref{Fig:bare_ABA}, the parabolic BLG-like bands are formed by atoms B$_1$, A$_2$, and B$_3$, which form the non-dimer interlayer pairs. The Dirac bands are formed by orbitals from the outermost graphene layers. 

\begin{figure*}[htb]
     \includegraphics[width=.8\textwidth]{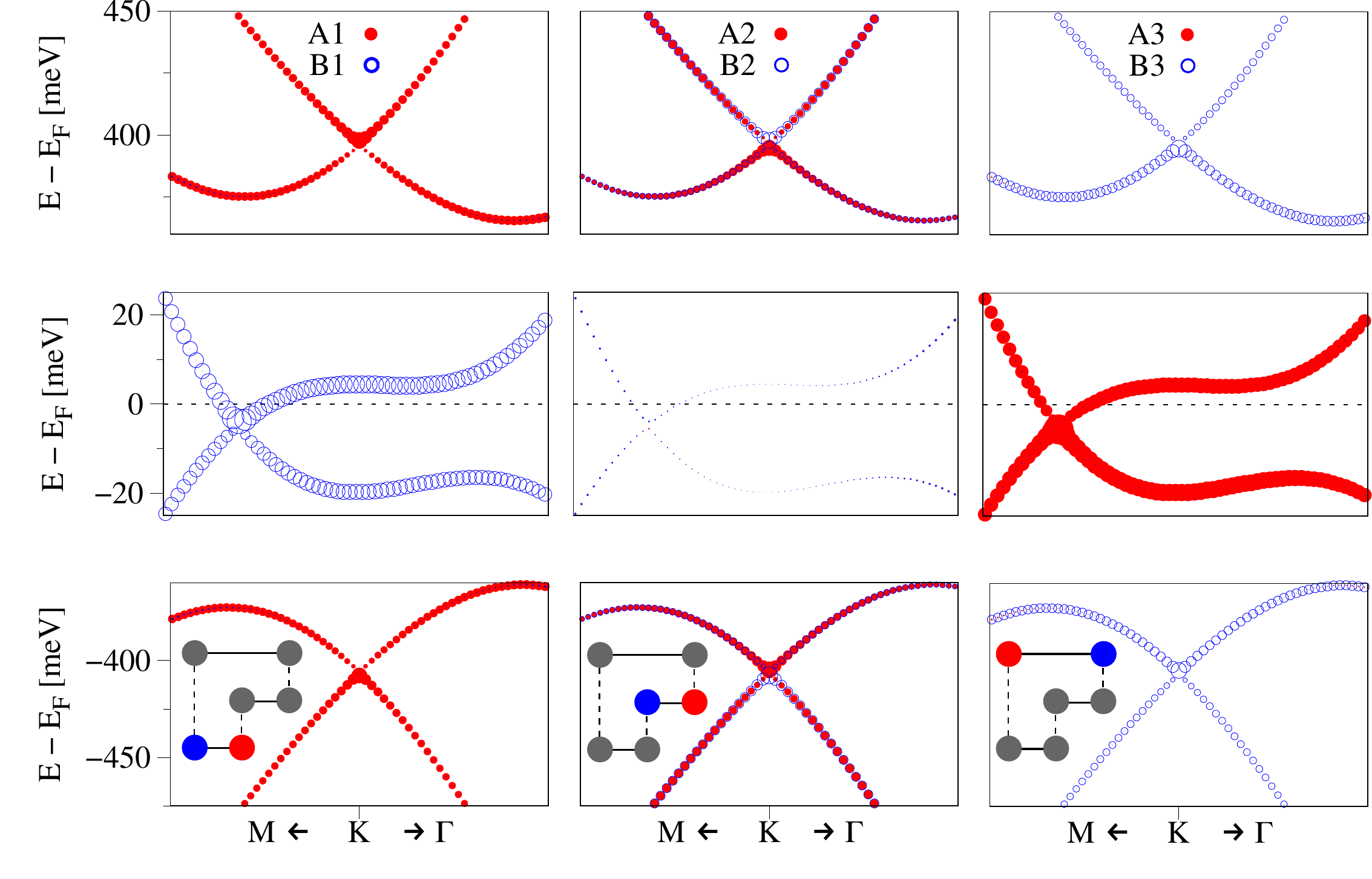}
     \caption{Zooms to the TLG bands in the vicinity of the $K$ point of the MoSe$_2$/ABC-TLG/WSe$_2$ heterostructure. From left to right, we project onto the sublattice atoms (A = red-filled circles, B = blue-open circles) of the bottom, middle, and top graphene layers. The sublattice atoms are indicated in the sketch of the ABC TLG geometry in the lower panels.
     }\label{Fig:pseudospin_ABC}
    \end{figure*}

In Fig.~\ref{Fig:pseudospin_ABC}, we show the pseudospin character of the ABC TLG bands of the TMDC encapsulated heterostructure. Consistent with Fig.~\ref{Fig:bare_ABC}, the flat low-energy bands are formed by the outer-layer atoms B$_1$ and A$_3$, while high-energy bands are formed by all other atoms.

\bibliography{paper}

\begin{thebibliography}{139}
\bibitem [{dos()}]{dos}%
  \BibitemOpen
  \href@noop {} {}\bibinfo {note} {Calculating the DOS from DFT would require a
  dense sampling of the Brillouin zone and a huge number of $k$ points, which
  is computationally demanding, given also the large number of atoms in the
  encapsulated TLG supercells. We circumvent this problem by employing the
  model Hamiltonian, which very accurately reproduces the low-energy
  dispersions. For the model DOS calculations, we employ a $k$-point grid of
  $5000 \times 5000$ in the vicinity ($\pm 0.1$ \AA$^{-1}$) of the $K$
  point.}\BibitemShut {Stop}%
\end{thebibliography}

\end{document}